%% file: paper.tex
\documentclass[sigconf,nonacm]{acmart}

\input{header}

\def\tool{\textsc{RustAssistant}\xspace}

\begin{document}

\title{Fixing Rust Compilation Errors using LLMs}

\author{Pantazis Deligiannis}
\affiliation{%
  \institution{Microsoft Research}
  \city{Redmond, WA}
  \country{USA}}
\email{pdeligia@microsoft.com}

\author{Akash Lal}
\affiliation{%
  \institution{Microsoft Research}
  \city{Bengaluru}
  \country{India}}
\email{akashl@microsoft.com}

\author{Nikita Mehrotra}
\affiliation{%
  \institution{Microsoft Research}
  \city{Bengaluru}
  \country{India}}
\email{t-nmehrotra@microsoft.com}

\author{Aseem Rastogi}
\affiliation{%
  \institution{Microsoft Research}
  \city{Bengaluru}
  \country{India}}
\email{aseemr@microsoft.com}

\input{abstract}

\maketitle

\input{introduction}
\input{overview}
\input{implementation}
\input{dataset}
\input{evaluation}
\input{threats}
\input{related}
\input{conclusions}

\bibliographystyle{ACM-Reference-Format}
\bibliography{references}

\end{document}

%% file: header.tex
\usepackage{url}
\usepackage{booktabs}
\usepackage{multirow}
\usepackage{pifont}
\usepackage{graphicx}
\usepackage{textcomp}
\usepackage{listings}
\usepackage{subcaption}
\usepackage{tcolorbox}
\usepackage{xspace}
\usepackage{xcolor}
\usepackage{algpseudocode}
\usepackage{ulem}
\usepackage{enumitem}
\usepackage{algorithm}

\hypersetup{
    colorlinks = true,
    linkbordercolor = {blue}
}

\definecolor{codeblue}{RGB}{0,110,179}
\definecolor{codegreen}{rgb}{0,0.6,0}
\definecolor{codered}{RGB}{220,20,60}
\definecolor{codegray}{rgb}{0.5,0.5,0.5}
\definecolor{codepurple}{rgb}{0.58,0,0.82}
\definecolor{codegold}{RGB}{106,128,0}
\definecolor{bg_green}{RGB}{252,255,245}
\definecolor{bg_blue}{RGB}{0,51,102}

\tcbuselibrary{listings}
\tcbset{
  fonttitle=\footnotesize,
  center title,
  colback=bg_green,
  colframe=bg_blue,
  listing only,
  listing options={
    basicstyle=\tt\scriptsize,
    backgroundcolor=\color{bg_green},   
    commentstyle=\color{codegreen},
    keywordstyle=\color{magenta},
    stringstyle=\color{codepurple},
    numbers=none,
    breakatwhitespace=false,
    breaklines=true,
    keepspaces=true,
    showspaces=false,
    showstringspaces=false,
    showtabs=false,
    tabsize=4,
    aboveskip=0pt,
    belowskip=0pt
  },
  boxrule=1pt,
  arc=2mm,
  toptitle=-0.1em,
  bottomtitle=-0.2em,
  lefttitle=0.1em,
  righttitle=0.2em,
  top=1pt,
  bottom=1pt,
  left=-0.02em,
  right=-0.2em
}

\lstdefinelanguage{rust}{
  basicstyle=\small,
  morekeywords=[1]{
    abstract, as, async, await, become, box, break, const, continue, crate, do, dyn,
    else, enum, extern, false, final, fn, for, if, impl, let, loop, macro, match,
    mod, move, mut, override, priv, pub, ref, self, Self, static, struct, super,
    trait, true, try, typeof, union, unsafe, unsized, virtual, where, while,
    yield
  },
  morekeywords=[2]{
    ChangeLog, FixDescription, OriginalCode, FixedCode
  },
  morekeywords=[3]{},
  morekeywords=[4]{
    isize, usize, i32, u32, i64, u64, f32, f64
  },
  morestring=[b]",
  sensitive=true,%
  numbersep=5pt,
  numbers=left,
  columns=[l]fullflexible,
  texcl=true,
  mathescape=true,
  xleftmargin=10pt,
  identifierstyle={\sffamily},
  keywordstyle=[1]{\sffamily\color{blue}},
  keywordstyle=[2]{\sffamily\color{purple}},
  keywordstyle=[3]{\sffamily\color{codegold}},
  keywordstyle=[4]{\sffamily\color{teal}},
  rangeprefix=(*---\ ,
  includerangemarker=false,
  stringstyle=\ttfamily,
  showspaces=false,
  morecomment=[l]{//},
  morecomment=[n]{(*}{*)},
  commentstyle={\footnotesize\itshape\color{codegreen}},
  literate={^-^}{{\tt\textcolor{codered}{- }}}{1}
           {^+^}{{\tt\textcolor{codegreen}{+ }}}{1},
  moredelim=[is][\color{purple}]{<@}{@>},
  breaklines=true
}

\newcommand{\cmark}{\ding{51}}
\newcommand{\xmark}{\ding{55}}

\let\ls\lstinline

\newcommand{\link}[2]{\href{#1}{\uline{#2}}}

%% file: abstract.tex
\begin{abstract}
The Rust programming language, with its safety guarantees, has established itself as a
viable choice for low-level systems programming language over the traditional, unsafe
alternatives like C/C++. These guarantees come from a strong ownership-based type system, as
well as  primitive support for features like closures, pattern matching, etc., that make 
the code more concise and amenable to reasoning. These unique Rust features also
pose a steep learning curve for programmers. 

This paper presents a tool called \tool that leverages the emergent capabilities
of Large Language Models (LLMs) to automatically suggest fixes for Rust
compilation errors. \tool uses a careful combination of prompting techniques as
well as iteration with an LLM to deliver high accuracy of fixes. \tool is able
to achieve an impressive peak accuracy of roughly $74\%$ on real-world
compilation errors in popular open-source Rust repositories.
We plan to release our dataset of Rust compilation errors to enable further research.
\end{abstract}

%% file: introduction.tex
\section{Introduction}
\label{sec:introduction}

The emergence of Large Language Models (LLMs) and their code comprehension
capabilities is disrupting the way we build, maintain, and deploy software
systems. LLMs are rapidly becoming an integral part of the workflow, starting
from software development~\cite{copilot}, testing~\cite{copilotx},
repair and debugging~\cite{chatdbg,DBLP:journals/corr/abs-2208-11640,DBLP:journals/corr/abs-2303-07263,DBLP:conf/sp/PearceTAKD23,DBLP:journals/corr/abs-2112-02125,DBLP:conf/icse-apr/PrennerBR22,DBLP:conf/icse/XiaWZ23,DBLP:conf/icse/FanGMRT23},
to how the users interact with the software~\cite{bingai,officecopilot}.

LLMs are large models, that have been trained on vast amount of public data,
including open-source
code~\cite{gpt35,DBLP:journals/corr/abs-2303-08774,DBLP:journals/corr/abs-2302-13971,DBLP:journals/corr/abs-2307-09288,DBLP:journals/corr/abs-2305-10403}.
The promise of LLMs is
that, without any need for fine-tuning the model for the task at hand, one can
interact with them using natural language \textit{prompts} and get them to follow
the instructions to accomplish the task. Using prompt engineering to study and
harness the novel emergent behaviors of LLMs has become an active area of
research~\cite{saparov2023language,DBLP:conf/icse/NashidSM23}.

In this paper, we consider the task of fixing Rust~\cite{rustlang} compilation
errors using LLMs. Rust, with its safety guarantees, has established itself as a
viable choice for low-level systems programming language over the traditional, unsafe
alternatives like C/C++. Rust enjoys strong support from both the open-source
community (e.g., support for Rust in the Linux kernel~\cite{rustlinux},
voted as the most loved language in the Stack Overflow survey~\cite{sosurvey})
and technology companies alike~\cite{rustwindows,amazonrust}.

The Rust typechecker, with a novel \textit{borrow checker} at its core,
ensures that the Rust programs are free of memory-safety errors and data races
that have plagued low-level systems for decades\footnote{A Microsoft study found that
$\sim70\%$ of the vulnerabilities Microsoft assigns a CVE each year continue to be
memory safety issues~\cite{mssurveymemory}.}.
The main idea is that every value is used \textit{linearly} and
has exactly one owner. A value may be \textit{borrowed} using references---there may
exist multiple immutable borrows or a single, exclusive mutable borrow for a value
at any point in the program. In addition, Rust has primitive support for features like
closures, pattern matching, etc., that make the code more concise and amenable
to reasoning.

\begin{figure}
  \begin{lstlisting}[
      language=rust,
      basicstyle=\footnotesize,
      frame=single,
      numberstyle=\tiny,
      numbers=left,
      firstnumber=15
  ]
struct Foo { map: RwLock$<$HashMap$<$String, Bar$>$$>$ }
  
impl Foo {
  pub fn get(&self, key: String) $\rightarrow$ &Bar {
    self.map.write().unwrap().entry(key).or_insert(Bar::new())
  }
}
\end{lstlisting}
\vspace{-1em}
\caption{Snippet from a Stack Overflow question}
\label{fig:overview-example}
\vspace{-1em}
\end{figure}

However, this also means that there is a steep learning curve for programmers
coming to Rust from the traditional C/C++ background. Although Rust tooling
(compiler error messages, IDE support~\cite{rustanalyzer}) is well-designed to help programmers
understand and fix the compilation errors, they can still be intimidating for
Rust beginners. A recent survey by the Rust team~\cite{rustsurvey}, for example,
reports that $83\%$ of the responders who adopted Rust at work found it to be
\textit{challenging}. Though adopting a new language and its ecosystem is always
challenging, $27\%$ of the responders also say
that using Rust is \textit{at times a struggle}.

Consider, for example,
Figure~\ref{fig:overview-example} showing snippet from a Stack Overflow
question~\cite{rust-so-question} 
about the following error that the Rust compiler emits on this code:

\begin{lstlisting}[language=rust,basicstyle=\footnotesize,numbers=none,frame=single,keepspaces=true]
  error[E0515]: cannot return value referencing temporary value
  --> src/example.rs:21:4
       |
  21 |     self.map.write().unwrap().entry(key).or_insert(Bar::new())
       |     -------------------^^^^^^^^^^^^^^^^^^^^^^^^^
       |     |
       |     returns a value referencing data owned by the current function
       |     temporary value created here
\end{lstlisting}

Fixing this error is not straightforward. For non-experts, making sense of the Rust borrow checking rules,
especially with mutex and concurrency can be daunting. And even if the solution is conceptually
clear to the experts, they still need to know about the Rust libraries to use for the
fix (we get back to the example in Section~\ref{sec:overview}).

\paragraph{Contributions}
We present \tool, an LLM-based tool for automatically fixing Rust compilation errors.
\tool is a command-line tool that given a filesystem path of a Rust development,
potentially containing build errors, iterates with an LLM 
to fix those errors and generate patches.
\tool specializes prompt construction for the purpose of fixing compilation errors, and
uses a novel changelog format in order to harness
the LLM capabilities (Section~\ref{sec:implementation}).

To evaluate \tool, we built a dataset of Rust compilation errors collected
from three different sources: (a) 270 micro-benchmarks that we have written
ourselves, covering 270/506 official Rust error codes~\cite{rust-error-codes},
(b) 50 Rust programs collected from Stack Overflow questions, and
(c) 182 GitHub commits with compilation errors from the top-100 Rust crates from
\texttt{\small{crates.io}} (Section~\ref{sec:dataset}).

Using \tool and our dataset, we systematically study and evaluate capabilities of
two LLMs, GPT-3.5~\cite{gpt35} and GPT-4~\cite{DBLP:journals/corr/abs-2303-08774},
for fixing Rust compilation
errors. With GPT-4, we find that \tool is able to fix $92.59\%$ of the micro-benchmarks,
$72\%$ of the Stack Overflow programs, and $73.63\%$ of the GitHub commits. We also
find that GPT-4 performs better than GPT-3.5 in the task. We report on several
ablation studies, to investigate the impact of algorithmic and
prompting variations~(Section~\ref{sec:evaluation}). 

We also demonstrate the
generalizability of \tool on errors beyond compilation errors. We show that
\tool, with minor prompt modifications, can handle linter errors generated by
Rust Clippy, one of the most popular static analysis tool for Rust
\cite{rust-clippy}. \tool achieves an accuracy of $75\%$ on the top-10 Rust crates
for fixing Clippy-reported errors, which is almost $2.4$ times better fix rate 
than Clippy's own auto-fix feature.

We plan to open-source both our dataset as well as the implementation of \tool.

%% file: overview.tex
\section{Overview}
\label{sec:overview}

In this section we describe the scope of our work and walk through the \tool
pipeline using the example from Section~\ref{sec:introduction}.

\subsection{Scope}

Our goal is to build a toolchain and systematically evaluate the capabilities of LLMs for
generating fixes for Rust compilation errors. These fixes must pass the
compiler and must also retain the \textit{intended semantics} of the code. The
former is an objective criterion while the latter is subjective. Since the code that we start with
is not even well-typed, let alone have a well-defined semantics, one requires
external judgement to assess the quality of a fix. In the evaluation of \tool
(Section~\ref{sec:evaluation}), we either rely on test cases 
(fix must build and pass the test) or a comparison with the actual fix made by a
developer to establish quality.

We focus on fixing code-related issues and, thus, consider editing only the
Rust source files (i.e., files with \texttt{\small{.rs}} extension).
Errors that require changing a configuration
(e.g., adding a package to a \texttt{\small{.toml} }file) are currently out of scope.
We also do not consider errors related to the use of the \ls{unsafe} keyword, which
provides an escape hatch from the typechecker.

\subsection{Overview of the \tool pipeline}

The code snippet in Figure~\ref{fig:overview-example} contains \ls{struct Foo}, which
maintains a \ls{HashMap} mapping \ls{String} keys to \ls{Bar} values. The hashmap is
concurrency-protected with a \ls{RwLock}, a locking mechanism in Rust that allows multiple
readers but at-most one writer at a time. The programmer's intent in the function
\ls{get} is to return a reference to the value mapped to \ls{key} in the hashmap.

The implementation of \ls{get} first calls \ls{RwLock::write}, a blocking call that
returns a guarded, exclusive, write access to the object protected by the lock, in this
case the hashmap. The write access is released
as the guard goes out-of-scope, e.g., when the function call returns.
The function then proceeds to read the value of \ls{key}, and return a reference to
the read value. When this code is compiled with Rust, the compiler complains with the
error shown in Section~\ref{sec:introduction}.

Rust maintains a list of all the error codes that can be emitted by the
compiler~\cite{rust-error-codes}. Here the
the error code is \link{https://doc.rust-lang.org/error_codes/E0515.html}{E0515} on line \ls{21}, meaning that the function is trying to return
reference to a local variable. The error comes from the Rust borrow checker.
The returned reference, which is derived from the mutex owned hashmap, escapes the function scope,
and therefore, outlives the mutex guard lifetime---a violation of the borrow checking rules.

Let's see how \tool fixes the code. \tool first invokes the Rust compiler on
the input code and collects the error message. It then
feeds the code and the compilation error, along with instructions in a prompt, to
the LLM. \tool is parametric in the choice of LLM; we show the interactions with GPT-4 in this
section.  Figure~\ref{fig:overview-example-fix1} shows the output of GPT-4. The output
contains the suggested fix in text, and a form of \textit{changelog}. \tool prompts the
LLM to use this specific format; we explain the LLM prompt
and the output format in detail in the next section.

The suggested fix is to change the return type of \ls{get} to \ls{Arc<Bar>}, and also
insert \ls{Arc<Bar>} in the hashmap.
\ls{Arc} in Rust is an atomically reference-counted, thread-safe pointer.
With this change, the \ls{get}
function can return a copy of the value, an \ls{Arc} pointer that points to the same
heap location as the value in the map (\ls{clone} creates the copy).

\begin{figure}
\begin{lstlisting}[language=rust,basicstyle=\footnotesize,frame=single,numbers=none,keepspaces=true]
ChangeLog:1@src/example.rs
FixDescription: Change the return type of the 'get' method to return an
Arc<Bar> and wrap the Bar in an Arc when inserting it into the HashMap.
<@OriginalCode@>@19-23:
[19] impl Foo {
[20]   pub fn get(&self, key: String) $\rightarrow$ &Bar {
[21]     self.map.write().unwrap().entry(key).or_insert(Bar::new())
[22]   }
[23] }
<@FixedCode@>@19-24:
[19] impl Foo {
[20]   pub fn get(&self, key: String) $\rightarrow$ std::sync::Arc<Bar> {
[21]     self.map.write().unwrap().entry(key).or_insert_with(
[22]       || std::sync::Arc::new(Bar::new())).clone()
[23]   }
[24] }
\end{lstlisting}
\vspace{-1em}
\caption{Output of \tool.}
\label{fig:overview-example-fix1}
\vspace{-1em}
\end{figure}

\begin{figure}
\begin{lstlisting}[language=rust,basicstyle=\footnotesize,numbers=none,frame=single,keepspaces=true]
   error[E0308]: mismatched types
   --> src/example.rs:22:6
        |
   22 |       || std::sync::Arc::new(Bar::new())).clone()
        |         ^^^^^^^^^^^^^^^^^^^^^^^^^
        |         expected struct `Bar`, found struct `Arc
\end{lstlisting}
\vspace{-1em}
\caption{Error after the first fix suggested by \tool}
\label{fig:overview-example-error-2}
\vspace{-1em}
\end{figure}

\tool parses this LLM output, applies the suggested patch to the code, and
compiles the program again.
This time, the Rust compiler complains with the error shown in Figure~\ref{fig:overview-example-error-2}.

The error is about the mismatch between the declared type of
hashmap, mapping \ls{String} keys to \ls{Bar} values, and the usage of it as a map
from \ls{String} keys to \ls{Arc<Bar>} values---indeed, the previous patch did not fix the
declaration of the hashmap.

\tool sends the code and the error to the LLM again. In this instance, GPT-4
responds with the following fix, correctly suggesting to change the type of \ls{map}.

\begin{lstlisting}[language=rust,basicstyle=\footnotesize,numbers=none,frame=single,keepspaces=true]
ChangeLog:1@src/example.rs
FixDescription: Change the type of values stored in the HashMap to Arc<Bar>.
<@OriginalCode@>@16-16:
[16]   map: RwLock<HashMap<String, Bar>>
<@FixedCode@>@16-16:
[16]   map: RwLock<HashMap<String, std::sync::Arc<Bar>>
\end{lstlisting}

\tool parses the output, applies the patch to the code, and invokes the Rust
compiler again. This time the compiler succeeds and the tool returns.
Using \texttt{Arc} is also the accepted Stack Overflow
answer for this question~\cite{rust-so-question}.

%% file: implementation.tex
\section{\tool Implementation}
\label{sec:implementation}

\begin{algorithm}
\caption{The \tool algorithm.}
\label{fig:algorithm}
\begin{algorithmic}[1]
  \Require $m$: LLM, $N$: Number of completions
  \Require $project$: Rust project
  \State $errs\gets \mathsf{check}(project)$
  \While{$errs \not= \emptyset$}
  \State $e \gets \mathsf{choose\_any}(errs)$
  \State $g\gets \{e\}$
  \State $snap\gets project$
  \While{$g \not= \emptyset$}
  \State $e'\gets \mathsf{choose\_any}(g)$
  \State $p\gets \mathsf{instantiate\_prompt}(e')$
  \State $n\gets \mathsf{invoke\_llm}(m, p, N)$
  \State $c\gets \mathsf{best\_completion}(project, n)$
  \State $project\gets \mathsf{apply\_patch}(project, c)$
  \State $g\gets \mathsf{check}(project) - errs$ 
  \If{$\mathsf{giveup()}$}
    \State $project \gets snap$
    \State \textbf{break}
  \EndIf
  \EndWhile
  \State $errs\gets \mathsf{check}(project)$
  \EndWhile
\end{algorithmic}
\end{algorithm}

\tool is a command-line tool that takes as input the filesystem path to
a Rust project, potentially with errors. For instance, the project may have
compilation errors, reported by the Rust compiler, or linting errors
reported by a tool like Rust Clippy \cite{rust-clippy}. 
We keep the notion of the underlying checker (Rust compiler or clippy) and
the errors (build errors or lint errors) abstract in this section.
\tool parses the project to build an in-memory
index of the Rust source files. The index allows \tool to retrieve the contents
of the files, edit them, or even revert them to a previous state.

\tool must handle the complexities of fixing errors in real-world scenarios.
Source files can be large relative to the LLM prompt sizes that were available to us
(maximum of $32$K tokens, for GPT-4), and most of the code in a file might not
be relevant to a reported error any way. \tool, therefore, performs \textit{localization} for each error to
identify relevant parts of the source code and presents only those parts to the
LLM, i.e., a single prompt may contain multiple code snippets. This
implies that we need a way of parsing the LLM response to know which change
needs to be applied where. We tried a naive approach where we asked the LLM to
simply give us the revised code snippets. This approach did not work in our
real-world evaluation. As Section~\ref{sec:evaluation} will show
(Table~\ref{tab:result_rustcrate_ablation}),
the accuracy of \tool tanked below $10\%$, making the tool unusable. To account
for this, 
we define a simple, but effective, \textit{changelog} format that only captures
the \textit{changes} that need to be made to the given code snippets. We
describe this format in the prompt and instruct the model to follow it. 
\tool uses a lightweight parser to understand the changelog in the LLM's
response and can then easily apply the changes to the original source code.
This approach significantly increases \tool's accuracy, essentially because the
LLM stays focussed on the changes that it needs to make (more details in
Section~\ref{sec:evaluation}). This justifies why our prompt construction is
an important contribution. 

Algorithm~\ref{fig:algorithm} shows the \tool core algorithm. The algorithm
starts by invoking the checker on the project to gather the
initial list of errors, and starts fixing them one at a time (line 2).

\paragraph{\textbf{Inner loop for fixing an input error}}
The inner loop (lines $6-17$) iterates with the LLM with the goal of fixing a
single input error ($e$). 
During this iteration, the source files may change and those changes may
themselves induce additional errors. 
To accommodate this, we introduce an abstraction called an \textit{error group} as the
working unit of the \tool inner loop. An error group is a set of errors that 
\tool is currently trying to fix. An error group is
initialized with the input error (line 4) and may grow or
shrink within the loop. The loop terminates when either the error group becomes
empty, implying that the original error $e$ was fixed, 
or \tool gives up on the error group (line 13), in which case the project is restored to
its initial state at the beginning of the output loop.
We now explain the body of the inner loop. 

\newcommand{\prompttext}[1]{\texttt{\small{#1}}}

\paragraph{\textbf{Prompt construction (\textsf{instantiate\_prompt})}}
For each error ($e'$) in the current error group, \tool constructs 
a prompt $p$, shown in Figure~\ref{fig:prompt-template} (the headings are for
illustration purposes only), asking for a fix to the error.
The prompt is parameterized over error-specific content, using the \texttt{\small{`\{\}'}}
syntax.
The \prompttext{preamble} section instantiates the checker command that was used
(\prompttext{cmd}). The next section of the prompt contains the error and
its textual explanation. For the Rust compilation errors, the errors
are self-explanatory, so we use the text from the error message as its explanation.
For the Clippy lint errors, we use the Rust command
\texttt{\small{cargo clippy {-}{-}explain ERROR\_CODE}} to
fill-in the explanation. This is followed by code snippet(s) that \tool
deems necessary to present to the LLM for fixing the error. These are obtained
by first identifying source locations in the error. The Rust compiler, for
instance, not just points to the error location, but to related locations as well. 
In the example error message below, the location after \ls{note} is a related location:

\begin{lstlisting}[language=rust,basicstyle=\footnotesize,numbers=none,frame=single,keepspaces=true]
error[E0369]: binary operation `>=' cannot be applied to `Verbosity'
--> src/logger.rs:53:21
     |
53 |     if self.verbosity >= Verbosity::Exhaustive {
     |     ----------- ^^^ -------------
note: an implementation of `PartialOrd<_>' might be missing
--> src/logger.rs:16:1
     |
16 | pub enum Verbosity {
     | ^^^^^^^^^^^^^^^^ must implement `PartialOrd<_>'
help: consider annotating with `#[derive(PartialEq, PartialOrd)]'
     |
16 | #[derive(PartialEq, PartialOrd)]
\end{lstlisting}

\tool then extracts a window of $\pm$50 lines (configurable)
around each location, and adds these snippets to the
prompt. It also adds the line
number for each line of code as a prefix, which helps the LLM to
better identify the code lines in the prompt. In an initial attempt, we tried
only extracting code segments in a proper lexical scope (e.g., the entire body of
a function where a relevant line appears). This not only increase the complexity
of our tooling (because one needs to parse the Rust code and obtain an AST) but
we also found that LLMs are robust even to non-lexical scopes. We, hence,
decided in favor of keeping our tooling simple. 

\begin{figure}[hbt!]
\centering
\begin{subfigure}{\linewidth}
\centering
\begin{tcblisting}{title=\tool prompt template preamble}
You are given the below error from running '{cmd}' and Rust code
snippets from one or more '.rs' files related to this error.
\end{tcblisting}
\end{subfigure}
\begin{subfigure}{\linewidth}
\centering
\begin{tcblisting}{title=Prompt context with error information and code snippets}
{error} {error_explanation}
---
{code_snippets}
\end{tcblisting}
\end{subfigure}
\begin{subfigure}{\linewidth}
\centering
\begin{tcblisting}{title=Instructions for fixing the error}
Instructions: Fix the error on the above code snippets. Not every
snippet might require a fix or be relevant to the error, but take
into account the code in all above snippets as it could help you
derive the best possible fix. Assume that the snippets might not
be complete and could be missing lines above or below. Do not add
comments or code that is not necessary to fix the error. Do not
use unsafe or unstable features (through '#![feature(...)]'). For
your answer, return one or more ChangeLog groups, each containing
one or more fixes to the above code snippets. Each group must be
formatted with the below instructions.
\end{tcblisting}
\end{subfigure}
\begin{subfigure}{\linewidth}
\centering
\begin{tcblisting}{title=Instructions and examples for formatting the changelog output}
Format instructions: Each ChangeLog group must start with a
description of its included fixes. The group must then list one
or more pairs of (OriginalCode, FixedCode) code snippets. Each
OriginalCode snippet must list all consecutive original lines of
code that must be replaced (including a few lines before and
after the fixes), followed by the FixedCode snippet with all
consecutive fixed lines of code that must replace the original
lines of code (including the same few lines before and after the
changes). In each pair, the OriginalCode and FixedCode snippets
must start at the same source code line number N. Each listed
code line, in both the OriginalCode and FixedCode snippets, must
be prefixed with [N] that matches the line index N in the above
snippets, and then be prefixed with exactly the same whitespace
indentation as the original snippets above.
---
ChangeLog:1@<file>
FixDescription: <summary>.
OriginalCode@4-6:
[4] <white space> <original code line>
[5] <white space> <original code line>
[6] <white space> <original code line>
FixedCode@4-6:
[4] <white space> <fixed code line>
[5] <white space> <fixed code line>
[6] <white space> <fixed code line>
OriginalCode@9-10:
[9] <white space> <original code line>
[10] <white space> <original code line>
FixedCode@9-9:
[9] <white space> <fixed code line>
...
ChangeLog:K@<file>
FixDescription: <summary>.
OriginalCode@15-16:
[15] <white space> <original code line>
[16] <white space> <original code line>
FixedCode@15-17:
[15] <white space> <fixed code line>
[16] <white space> <fixed code line>
[17] <white space> <fixed code line>
OriginalCode@23-23:
[23] <white space> <original code line>
FixedCode@23-23:
[23] <white space> <fixed code line>
---
Answer:
\end{tcblisting}
\end{subfigure}
\vspace{-2em}
\caption{The \tool prompt template.}
\label{fig:prompt-template}
\vspace{-2em}
\end{figure}

The next section of the prompt (\prompttext{instructions}) are simple instructions that ask for a fix.
For instance, it instructs the model to avoid adding \textit{unsafe} code, in an
effort to keep the tool focused on generating good Rust code.

The final section of the prompt contains instructions to the LLM for
formatting the output, as a list of one or more \textit{change logs}. Each
changelog begins with an ID numbered starting with $1$  and the source file
to which it is applied (\texttt{ChangeLog} line
in Figure~\ref{fig:prompt-template}). Next is free-form description of the fix
(\texttt{FixDescription} line). This description is not parsed; it is only to
enable \textit{scratchpad} reasoning in the model
\cite{DBLP:journals/corr/abs-2112-00114}. Next is a repetition
of a part of the input code that was provided to the model (\texttt{OriginalCode}).
This part is \textit{defensive} because it is a repetition of the input; 
\tool rejects the changelog if the \texttt{OriginalCode} segment fails to match the actual original
code. Finally, the output must have the fixed code (\texttt{FixedCode}) that should replace all the
lines of the original code. If this segment is empty, for example, then it
implies that the corresponding original code segment should be deleted. There
are other defensive checks in the changelog format: each of \texttt{OriginalCode}
and \texttt{FixedCode} segments must mention the line number range; and this
number range repeats again in the code segment. All such checks act as a
guard; change logs are rejected when this information fails to match.

\paragraph{\textbf{LLM invocation (\textsf{invoke\_llm})}}
Once the prompt is instantiated, \tool invokes the LLM with the prompt
(line 9) asking for $N$ \textit{completions}, essentially, $N$ responses to the same prompt.
On receiving these completions, \tool ranks them and picks the 
best completion (line 10). To rank the completions, \tool applies all the changelogs
in a completion and counts the number of resulting errors reported by the
checker. The completion that results in the least number
of errors is ranked the highest. This is, in essence, a best-first search
strategy.

The best completion is applied to the project (line 11), the current error group is
updated (line 12) and \tool then continues with the inner loop. When the inner
loop completes, \tool updates the set of pending errors (line 18) because it is
possible that fixing one error group caused the errors to change. (As a detail,
errors that were previously given up, on line 13, are not tried again; but this
is omitted from the algorithm).

\tool uses a few heuristics to ensure termination of the inner and the outer
loops. First, it provides a configurable option (default 100) to limit the
maximum number of unique errors that an error group can have over its lifetime
in the inner loop. If this limit is reached, the inner loop gives up.
Second, if the set of errors in an error group does not change across
iterations of the inner loop, \tool considers it as not making progress and gives up on the error
group. The outer loop is bounded to run for as many iterations as the initial
number of errors obtained on line 1. (For the purpose of checking if two errors
are same, which is needed when performing set operations, 
\tool represents an error as the concatenation of its error code,
error message, and the file name, without any line numbers.)

%% file: dataset.tex
\section{Rust Error Dataset}
\label{sec:dataset}

We build a dataset of Rust compilation errors collected from three different 
sources, as well as linting errors from Clippy~\cite{rust-clippy} tool.

\par \noindent \textbf{Micro-benchmarks:} Rust offers a comprehensive catalog of
errors, indexed by error codes, that the Rust compiler may emit. The catalog is accompanied by small programs
that trigger the specific error
codes~\cite{rust-error-codes}. To build our micro-benchmarks dataset, we wrote
small Rust programs, one per error code,
designed specially to trigger that error code. Although we wrote
these programs ourselves, we used the snippets in the Rust catalog as a reference.

We consider 270 error codes out of a total of 506. We exclude error
codes that are no longer relevant in the latest version of the Rust compiler
(1.67.1). Additionally, we exclude all errors related to package use, build
configuration files, and foreign function interop, 
as well as error codes on the use of \ls{unsafe}; as mentioned in Section~\ref{sec:overview},
these errors are out-of-scope for us. We also create a unit test for each error code
that specifies the intended behavior of the program. We define passing this test as
the measure that the compilation error is fixed in a semantics-preserving way.

\par We further classified each of the errors codes into one of the six categories: 
Syntax, Type, Generics, Traits, Ownership, and
Lifetime. The primary objective of this benchmark is to
determine if \tool is more proficient at fixing certain types of errors
compared to others. 

\par \noindent \textbf{Stack Overflow (SO) code snippets:} 
Stack Overflow (SO) is a popular online community where programmers and developers
seek help for coding issues. 
We manually scrape SO to collect questions about Rust compilation errors.
To limit the effort, we concentrate on memory-safety and thread-safety
issues, two areas in which the Rust type system is stricter than C/C++.

\par To ensure that the questions are
relevant and substantial, we apply some filtering criteria. For example, we require
each question to have at least one answer and exclude cases that deem
trivial (e.g., the question is misclassified or
contains syntax errors unrelated to the intended category).
After applying these filtering criteria, we select the first 50 most relevant questions.

\par Code snippets in these questions are not always self-contained. We manually
add code and stubs to scope the compilation issue to only what was asked in the
corresponding SO question.
We manually add test cases also. Similar to the micro-benchmarks,
these test cases are designed to specify the intended semantics of the programs.

\par \noindent \textbf{Top-100 crates}: For a more comprehensive real-world
evaluation, we look at the GitHub repositories of the top-100 Rust crates (the most widely-used Rust
library packages) from \texttt{\small{crates.io}}~\cite{rust-crates-io}.
We examine the history of these repositories and identify 
commits that have compilation errors (we clone the commits and build them locally,
we also filter out commits where the errors are out-of-scope).
We found $182$ such commits. The benchmark then is to fix the commits so that they pass
the Rust typechecker. In our evaluation, we manually audit the fixes to check whether
they preserve the intended semantics (see RQ4 in Section~\ref{sec:evaluation}). 

To build a dataset of lint errors, we pick top-10 crates, and run \texttt{\small{rust-clippy}}
on the latest commit in the main branch of their corresponding GitHub repositories.
Clippy \cite{rust-clippy} is one of the most popular
open-source static analysis tool for Rust with roughly $10$K stars on
GitHub. It is designed to help
developers write idiomatic, efficient, and bug-free Rust code by providing a set
of predefined linting rules. Clippy also provides helpful messages and suggestions to guide developers in making
improvements to their code. Fixing Clippy errors tests the ability of \tool to
generalize beyond compilation errors. Our dataset has a total of $346$ Clippy
errors.

Clippy has multiple categories of checks \cite{rust-clippy}. 
For our dataset, we only consider
Pedantic, Complexity, and Style. The rest of the categories did not raise errors
in the top-10 crates. 
Additionally, there is a category called Nursery, but it consists of lints that
are not yet stable, so we exclude it from our consideration.
Pedantic refers to stylistic or convention violations, Complexity
refers to unnecessarily complex or convoluted code that hamper maintainability,
and Style covers various linting rules related to code style and best
practices, focusing on conventions such as naming, spacing, formatting,
and other stylistic aspects of the code.

%% file: evaluation.tex
\section{Evaluation}
\label{sec:evaluation}

The evaluation of \tool is designed to answer the following research questions:

\begin{enumerate}
    \item \textbf{RQ1:} To what extent is \tool successful in fixing Rust compilation
    errors?
    \item \textbf{RQ2:} How effective are different prompting strategies and
    algorithmic variations?
    \item \textbf{RQ3:} Can \tool generalize to fix errors reported by a Rust static analyzer?
    \item \textbf{RQ4:} How accurate are the fixes generated by \tool for real-world
    code repositories?
\end{enumerate}

\paragraph{LLM Configuration}
For creative and unconstrained responses, we set both the
\texttt{frequency\_penalty} and \texttt{presence\_penalty} parameters to 0. We also adopted a
deterministic approach by using \texttt{top\_p=1}, meaning that the most likely
token is selected at each generation step. To maintain the focus and consistency
of the outputs, we opt for a low temperature of $0.2$. While the maximum length of
the generated output is set to the default value of 800 tokens, in practice, our
experiments primarily involve returning concise changelog snippets, which are
significantly smaller in length.

\par We evaluate both GPT-3.5-turbo (300B parameters) \cite{gpt35} 
(which we call as GPT-3.5 in this paper) and GPT-4
\cite{DBLP:journals/corr/abs-2303-08774}; a comparison between them helps 
understand the effect of model scaling on \tool's accuracy. 
For both LLMs, we also vary the number of LLM completions (\#N) from
1 to 5 to provide insights into the optimal
balance between computation time and quality of the fixes.

\par \textbf{RQ1} \textit{``To what extent is \tool successful in fixing
Rust compilation errors?''}
For micro-benchmarks and SO code snippets, we report fix rate (Fix\%) as
the percentage of the programs that build and pass their associated unit test after
running \tool. A generated fix can fail for three reasons: (1) \textbf{Format
Errors}, where the generated changelog is not correctly formatted, or the
format is correct but the original code or lines do not match, leading to the
rejection of the changelog; (2) \textbf{Build Errors}, the fix when applied still
results in compilation errors; (3) \textbf{Failed Tests}, there are no
formatting or compilation errors, but the corresponding unit test fails.
For the top-100 crates benchmark, we report the number (percentage) of commits 
that successfully compile after running our tool. We also report the number
(percentage) of compilation errors fixed across all the commits.

\input{tables/error_codes.tex}

\begin{figure}[t]
\centering
\includegraphics[width=1\linewidth]{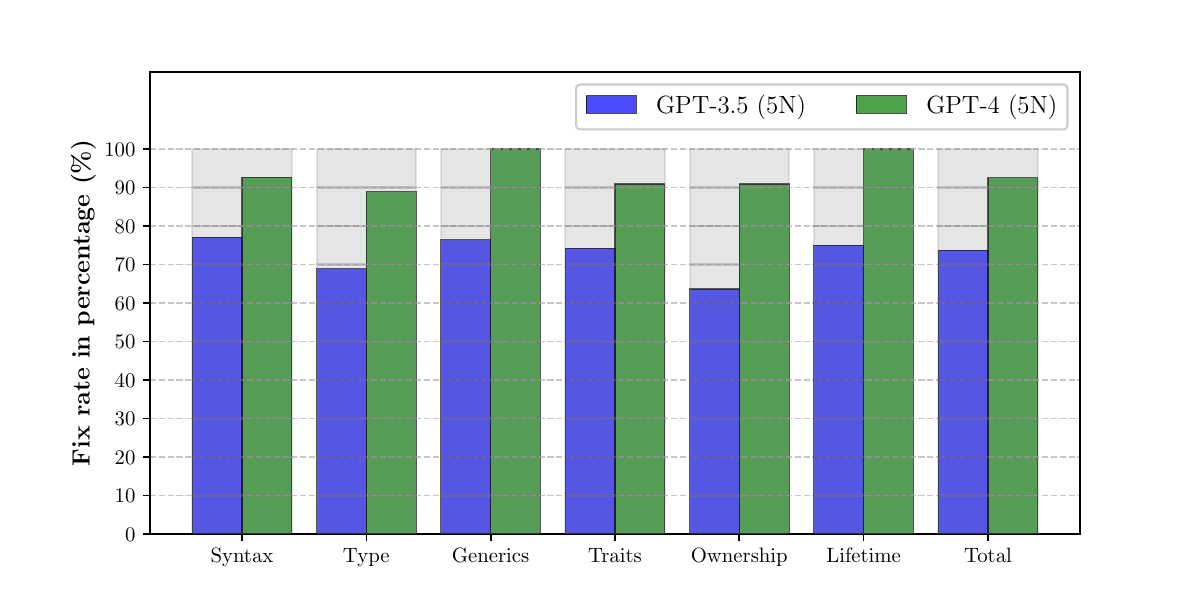}
\vspace{-2.5em}
\caption{Comparing the fix rate \% of GPT-3.5 and GPT-4 (both with $N=5$) across
error code categories.}
\label{fig:error_codes_plot}
\end{figure}

\input{tables/stack_overflow.tex}

\begin{figure}[t]
\centering
\includegraphics[width=1\linewidth]{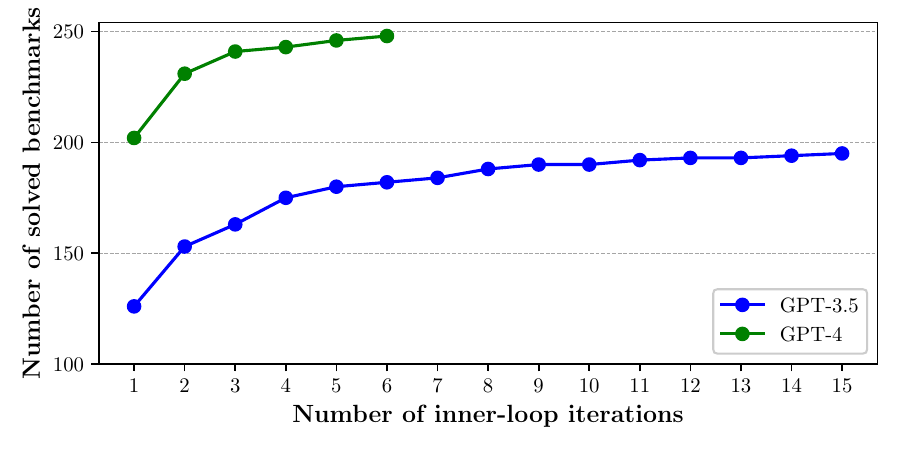}
\vspace{-2.5em}
\caption{Iterations required for micro-benchmarks.}
\label{fig:iterations_micro}
\end{figure}

\begin{figure}[t]
\centering
\includegraphics[width=1\linewidth]{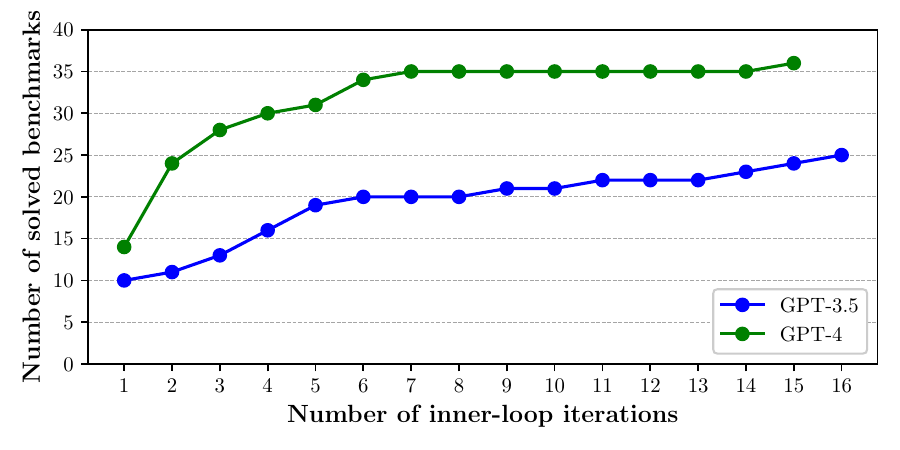}
\vspace{-2.5em}
\caption{Iterations required for SO benchmarks.}
\label{fig:iterations_so}
\end{figure}

\input{tables/top100.tex}

Table \ref{tab:result_microbenchmark} shows the performance of \tool on the
micro-benchmarks, achieving a peak accuracy of $92.59\%$. GPT-4's performance is
significantly better than GPT-3.5, so model scaling helps. 
Increasing the number of completions helps, but only minimally for GPT-4,
potentially because its fix rate is already very high with $N=1$.
We also notice that GPT-3.5 often fails to follow our formatting
requirement in its output ($59$ failures with $N=1$) or fails to satisfy the
compiler ($56$ failures with $N=1$). GPT-4 outshines GPT-3.5 on these aspects.
Interestingly, sometimes it produces a fix that fails the corresponding unit test
($13$ test failures with $N=1$); the following is an example:

\begin{lstlisting}[
    language=rust,
    basicstyle=\footnotesize,
    frame=single,
    numbers=none
  ]
pub fn get_value() $\rightarrow$ f64 {
  let mut val: f64 = 7.0;
  val <<= 2.0;
  val
}
\end{lstlisting}

This code fails to build because it uses the bitwise
operator \ls{<<=} on a floating point value (error code
\link{https://doc.rust-lang.org/error_codes/E0368.html}{E0368}). GPT-4
proposes a fix to change the operator to multiplication, but it keeps the
unit as \ls{2.0}, instead of changing it to \ls{4.0}. An alternate fix
could have been to use \ls{u32} instead of \ls{f64} in the example.
Expectedly, the GPT-4 fix fails our unit test.

Figure~\ref{fig:error_codes_plot} shows the fix rate across different error code
categories (outlined in Section~\ref{sec:dataset}). The performance of the models is fairly consistent
across categories,
suggesting that there are no specific error codes that are easier or harder for the
models to fix.

Table \ref{tab:result_so} presents results on SO benchmarks.
Overall trends, with respect to the two models and the number of completions, 
are similar to Table \ref{tab:result_microbenchmark}. However, the
Fix percentages are consistently lower. \tool is able to achieve a peak fix percentage of
$74\%$ demonstrating that these benchmarks are harder. Indeed as the code snippet in
Section~\ref{sec:introduction} shows, these examples use non-trivial Rust concepts.

Across the micro-benchmarks and SO benchmarks, we manually investigate the reasons for
failure. In some cases, the model suggests a correct partial fix but then
does not
follow up with the additional fixes required. In other cases, it gets stuck in 
a loop where it proposes a fix and undoes it in the next iteration, causing
\tool to give up. In a few cases, the model tries to import a package that it
needs, but \tool is not prepared to edit the \texttt{\small{.toml}} project file for
actually doing the import. It is possible that further refinement of the LLM prompt
can fix such issues; we leave it for future work.

For the cases where \tool produced a correct fix, 
Figures~\ref{fig:iterations_micro} and \ref{fig:iterations_so} show the number
of iterations of the inner-loop required for micro-benchmarks and SO benchmarks, respectively.
GPT-4 requires at most $6$ iterations for micro-benchmarks, but up to $15$
iterations in the harder SO benchmarks. GPT-3.5 typically requires much higher
number of iterations. In several cases, the iterations were indeed required; for
instance, if the type of a function parameter is changed (by, say, adding the \ls{mut}
qualifier) then the function call sites also need to change, etc.

Table \ref{tab:result_rustcrate} presents results on the top-100 crates benchmark. 
\tool is able to achieve an impressive peak accuracy of $91.46\%$ in terms of
fixing errors, matching what is also observed in the micro-benchmarks. When we
consider the ability to fix all errors in a commit, the fix rate is lower, but still
impressive at $73.63\%$, i.e., roughly three-fourths of the commits could have been
automatically fixed!

\par \textbf{RQ2} ``How effective are different prompting strategies and
algorithmic variations?'' We perform
an ablation study by permuting between different prompting and algorithmic
variations, in order to identify the most effective features of \tool.
We consider five prompt variants, which differ in the way \tool asks LLMs
to output the fixes, i.e. the output formatting instructions.

\begin{enumerate}
    \item \textbf{P0 (Basic)}: This variant serves as the baseline. It does not
      have the changelog section; it instead asks for the complete revised snippets. 
    \item \textbf{P1 (ChangeLog-basic)}: This uses the changelog, but only the
      \texttt{FixedCode} section, without the line number prefixes.
    \item \textbf{P2 (Line prefixes)}: In addition to P1, we require line
      number prefixes in front of code snippets. 
    \item \textbf{P3 (Localization)}: In addition to P2, we require the
      original code section. 
    \item \textbf{P4 (Description first)}: This is the full prompt of
      Figure~\ref{fig:prompt-template}, i.e., P3 with the
      \texttt{\small{FixDescription}} section. 
\end{enumerate}

\input{tables/top100ablation.tex}

For algorithmic ablations, we vary the number of completions (\#N) to either $1$ or
$5$ (already reported for RQ1), and we turn off error grouping. Without
error grouping, the \tool algorithm has a single loop that attempts to fix one error at a time
from the current bag of errors. 

Table \ref{tab:result_rustcrate_ablation} shows the results on
the top-100 crates benchmark. We see that P0 results in very poor performance
(compare its first two rows with those of Table \ref{tab:result_rustcrate}). We found that
the model, when returning the fixed code snippet would get tempted in making code
changes that were unrelated to the task of fixing the compiler error. 
This justifies the need for investing in changelog format to keep the model
focused on the fix. 

Table \ref{tab:result_rustcrate_ablation} also shows that turning off
grouping significantly drops the error fix rate (compare the last two rows of Table
\ref{tab:result_rustcrate_ablation} with the first two rows of Table
\ref{tab:result_rustcrate}). Without grouping, \tool would fix errors in a
random order, which increased the chances of it getting stuck with an error that
it could not fix, leading to a ever-increasing blow-up of code changes and
resulting errors. Error grouping helps in detecting such cases, allowing
\tool to gracefully give up on them, and then move on to the other errors in the
project. This justifies the importance of error grouping.

\input{tables/prompt_variants.tex}
Table \ref{tab:pvariants} presents the results for prompt ablations with GPT-3.5
(we skip GPT-4 due to limited capacity with the model). It demonstrates that the addition of each new
feature to the changelog format raises accuracy by a significant margin. The
basic format (P1) only provides roughly $10\%$ accuracy. We saw that P1 response would trip 
most on the formatting of its output, an important requirement in order to handle large
code bases. The number of formatting errors reduce significantly as the
changelog format is improved. It is interesting that the simple act of
describing the fix (going from P3 to P4), helps the model accuracy
significantly.

\par \textbf{RQ3} ``Can \tool generalize to fix errors reported by a Rust
static analyzer?''
Fixing Clippy errors tests the ability of \tool to
generalize beyond compilation errors.
Clippy also comes with an auto-fix option that is based on
pattern-matching. We use it as a baseline for comparison. 

\input{tables/linter.tex}
Table \ref{tab:result_linter} present the results on fixing Clippy errors. \tool
is able to fix $2.4$x more errors than Clippy's own auto-fix option, achieving
a peak accuracy of nearly $75\%$. The accuracy on Complexity and Style
categories exceeds $90\%$. 

\input{tables/manualevaluation.tex}

\par \textbf{RQ4} ``How accurate are the fixes generated by \tool for real-world
code repositories?''
To answer RQ4, we qualitatively examine the fixes generated by \tool on all the
134 fixed commits from the top-100 crates benchmark.
As shown in Table~\ref{tab:manual_evaluation}, we categorize the commits into
4 categories. The categories are defined in term of individual error fixes,
and a commit is
classified into a category if all its fixes\footnote{Note that a commit can have multiple errors
and hence multiple fixes.} belong to the category or to
the categories above it. For example, a commit is classified as
\textit{Non-matching, same runtime behavior} if all its fixes are either
\textit{Unambiguous}, or \textit{Matching}, or \textit{Non-matching, same runtime behavior}.

Unambiguous fixes classify fixes for errors like syntax errors
(e.g., missing \ls{;} or braces),
missing instantiations for generics type parameters,
type-incorrect format string specifiers, etc. Since there is mostly a unique way to fix
these errors, and the \tool fix passes the Rust typechecker, we consider it good.

Matching fixes classify the fixes where
the fix matches the developer fix in the repository---we checked this by
comparing the patched code from \tool with the corresponding code from the
latest commit in the main branch of the repository.
This category contains non-trivial examples that provide some evidence that
the LLMs have learnt common Rust idioms.
For example, a common pattern to destruct
a pointer in Rust is to wrap it using \ls{Box::from_raw(ptr)}, and let the \ls{Box}
destructor call the destructor for the pointer. In a few instances in our benchmark,
\ls{void} returning functions had \ls{Box::from_raw(ptr)} as the last statement,
which the Rust compiler complains about as unused value. \tool fixes these errors
by rewriting it as \ls{drop(Box::from_raw(ptr))}, which matches the
actual fix in the repository.

Similarly, there are examples where the Rust compiler complains about a function
modifying an argument through an immutable reference. \tool fixed such cases by
changing the signature of the function to demand \ls{&mut} references.
Another class of errors in this category are deprecation warnings/errors, e.g.:

\begin{lstlisting}[language=rust,basicstyle=\footnotesize,numbers=none,frame=single]
error: use of deprecated associated function
`core::sync::atomic::AtomicPtr::<T>::compare_and_swap`:
Use `compare_exchange` or `compare_exchange_weak` instead
--> src/bytes.rs:1002:23
\end{lstlisting}

The fix produced by \tool follows the suggestion in the error, and
comes up with the following patch:

\begin{lstlisting}[language=rust,basicstyle=\footnotesize,numbers=none,frame=single]
^-^    let actual = atom.compare_and_swap(ptr as _, shared as _, Ordering::AcqRel);
^+^    let actual = atom.compare_exchange(ptr as _, shared as _, Ordering::AcqRel,
                                    Ordering::Acquire).unwrap_or_else(|x| x);
\end{lstlisting}

The third category, non-matching but same runtime behavior,
classifies the fixes where the fix produced by \tool
does not match the fix in the repository, but to the best-of-our estimation,
the fix does not alter the runtime behavior of the program (recall that the fix passes
the Rust typechecker). This is an interesting category where the LLM produces
reasonable fixes that do not match the programmer intent.

One example in this category is the following error:

\begin{lstlisting}[language=rust,basicstyle=\footnotesize,numbers=none,frame=single]
error[E0015]: cannot call non-const fn `ArrayString::<CAP>::capacity`
in constant functions
\end{lstlisting}

The error is in a function defined as \ls{const}:

\begin{lstlisting}[language=rust,basicstyle=\footnotesize,numbers=none,frame=single]
pub const fn remaining_capacity(&self) $\rightarrow$ usize
\end{lstlisting}

The compiler complains that the function calls another function \ls{capacity},
passing it the \ls{self} argument, but \ls{capacity} is a non-\ls{const} function.
LLM chose to fix this error by removing the \ls{const} qualifier from
\ls{remaining_capacity}, whereas the programmer fixed it
by adding \ls{const} qualifier to \ls{capacity}. We found similar instances
related to other qualifiers such as \ls{mut} and \ls{public}.

Another example in the category is the error in the following \ls{trait}
definition:

\begin{lstlisting}[language=rust,basicstyle=\footnotesize,numbers=none,frame=single]
pub trait MendSlice { fn mend(Self, Self) $\rightarrow$ Result<Self,(Self,Self)>; }
\end{lstlisting}

Rust compiler complains that it needs to statically know the size of \ls{Self},
and suggests bounding \ls{Self} with the \ls{Sized} trait:

\begin{lstlisting}[language=rust,basicstyle=\footnotesize,numbers=none,frame=single]
error[E0277]: the size for values of type `Self` cannot be known at compilation time
--> src/misc.rs:151:41
   |
151 |     fn mend(Self, Self) $\rightarrow$ Result<Self, (Self, Self)>;
   |                                         ^^^^^^^^^^^^ doesn't have a size known at compile-time
   |
   = note: only the last element of a tuple may have a dynamically sized type
help: consider further restricting `Self`
   |
151 |     fn mend(Self, Self) $\rightarrow$ Result<Self, (Self, Self)> where Self: Sized;
\end{lstlisting}

The \tool fix in this case was to bound the \ls{Self} argument at the level of
\ls{MendSlice} definition
\begin{lstlisting}[language=rust,basicstyle=\footnotesize,numbers=none,frame=single]
pub trait MendSlice: {Sized} { fn mend(Self, Self) $\rightarrow$ Result<Self,(Self,Self)>; }
\end{lstlisting}

whereas in the repository, the fix is what the error message suggested. This is an interesting
case since the LLM chose to ignore the suggestion in the error message.

The final category is for the fixes where the changes introduced by
the LLM, again to the best of our estimation, alter the runtime behavior of the
program. In a few of these cases, the LLM patch was blatantly wrong, e.g., it removed
some code, introduced alternate implementations of some unrelated functions, etc.
However, in some cases it was understandable that the LLM patch did not match the
developer. An example is as follows. Rust supports enumerated types and a
\ls{match} construct to
inspect the variant of the enum and execute different code based on the variant.
If the \ls{match} is not exhaustive, i.e. it doesn't mention all the variant cases,
the compiler raises an error. In a few cases of these errors in our benchmark,
LLM got it right
(e.g., a \ls{match} that is just converting enum variant name to a string), but when
the error occurred in the context of more involved \ls{match}, we found that LLMs
came up with a fix different from the developer.

\paragraph{Summary.} From our evaluation, we conclude that the pre-trained
LLMs seem to have internalized the knowledge of Rust syntax and commonly used Rust
idioms. They also follow the errors and come up with the relevant and intended fixes
in most cases. They do, however, require careful prompt construction, and the
iteration with a compiler was necessary especially for propagating 
changes across different parts of the code.

It is interesting future work to see if we can embed more Rust
idioms in the LLM prompt to match the developer intention, basically try to move the cases
in the third category of Table~\ref{tab:manual_evaluation} to the second. For example, we can instruct the LLM to first try
fixing the code with strict qualifiers (immutable, \ls{private}, \ls{const}). We would
also need to implement better contextualization to pass the relevant code snippets in
the prompt. 

%% file: tables/error_codes.tex
\begin{table}[t]
\centering
\setlength{\tabcolsep}{4pt}
\begin{tabular}{rc cc ccc}
& & & & \multicolumn{3}{c}{\texttt{\#}\textbf{Failures}}\\
\cmidrule(lr){5-7}

\textbf{Model}
& \texttt{\#}\textbf{N}
& \textbf{Fix}\texttt{\%}
& \texttt{\#}\textbf{Fixed}
& \textbf{Format}
& \textbf{Build}
& \textbf{Test}\\
\toprule

\multirow{2}*{\texttt{GPT-3.5}}
& $1$
& $52.96$\texttt{\%}
& $143$ / $270$
& $59$
& $56$
& $12$
\\

& $5$
& $73.70$\texttt{\%}
& $199$ / $270$
& $44$
& $10$
& $17$
\\

\midrule

\multirow{2}*{\texttt{GPT-4}}
& $1$
& $92.22$\texttt{\%}
& $249$ / $270$
& \xmark
& $8$
& $13$
\\

& $5$
& $\textbf{92.59}$\texttt{\%}
& $\textbf{250}$ \textbf{/} $\textbf{270}$
& $1$
& $5$
& $14$
\\

\bottomrule
\end{tabular}
\vspace{0.5em}
\caption{Evaluation on the 270 error code micro-benchmarks.}
\label{tab:result_microbenchmark}
\vspace{-3em}
\end{table}

%% file: tables/stack_overflow.tex
\begin{table}[t]
\centering
\setlength{\tabcolsep}{4pt}
\begin{tabular}{rc cc ccc}
& & & & \multicolumn{3}{c}{\texttt{\#}\textbf{Failures}}\\
\cmidrule(lr){5-7}

\textbf{Model}
& \texttt{\#}\textbf{N}
& \textbf{Fix}\texttt{\%}
& \texttt{\#}\textbf{Fixed}
& \textbf{Format}
& \textbf{Build}
& \textbf{Test}\\
\toprule

\multirow{2}*{\texttt{GPT-3.5}}
& $1$
& $24$\texttt{\%}
& $12$ / $50$
& $18$
& $18$
& $2$
\\

& $5$
& $50$\texttt{\%}
& $25$ / $50$
& $21$
& $2$
& $2$
\\

\midrule

\multirow{2}*{\texttt{GPT-4}}
& $1$
& $\textbf{74}$\texttt{\%}
& $\textbf{37}$ \textbf{/} $\textbf{50}$
& $1$
& $7$
& $5$
\\

& $5$
& $72$\texttt{\%}
& $36$ / $50$
& $3$
& $4$
& $7$
\\

\bottomrule
\end{tabular}
\vspace{0.5em}
\caption{Evaluation on 50 benchmarks from Stack Overflow.}
\label{tab:result_so}
\vspace{-2em}
\end{table}

%% file: tables/top100.tex
\begin{table}[t]
\centering
\setlength{\tabcolsep}{2pt}
\begin{tabular}{r ccc cc cc}

\textbf{Model}
& \texttt{\#}\textbf{N}
& \textbf{Commits}\texttt{\%}
& \texttt{\#}\textbf{Commits}
& \textbf{Errors}\texttt{\%}
& \texttt{\#}\textbf{Errors}\\
\toprule

\multirow{2}*{\texttt{GPT-3.5}}
& $1$
& $30.22$\texttt{\%}
& $55$ / $182$
& $44.76$\texttt{\%}
& $414$ / $925$
\\

& $5$
& $35.71$\texttt{\%}
& $65$ / $182$
& $55.03$\texttt{\%}
& $509$ / $925$
\\

\midrule

\multirow{2}*{\texttt{GPT-4}}
& $1$
& $54.40$\texttt{\%}
& $99$ / $182$
& $86.05$\texttt{\%}
& $796$ / $925$
\\

& $5$
& $\textbf{73.63}$\texttt{\%}
& $\textbf{134}$ \textbf{/} $\textbf{182}$
& $\textbf{91.46}$\texttt{\%}
& $\textbf{846}$ $\textbf{/} $$\textbf{925}$
\\

\bottomrule
\end{tabular}
\vspace{0.5em}
\caption{Evaluation on the top-100 Rust crates.}
\label{tab:result_rustcrate}
\vspace{-2em}
\end{table}

%% file: tables/top100ablation.tex
\begin{table}[t]
\centering
\setlength{\tabcolsep}{0.8pt}
\begin{tabular}{r ccc cc cc}

\textbf{Model}
& \textbf{Prompt}
& \textbf{G}
& \texttt{\#}\textbf{N}
& \textbf{Commits}\texttt{\%}
& \texttt{\#}\textbf{Commits}
& \textbf{Errors}\texttt{\%}
& \texttt{\#}\textbf{Errors}\\
\toprule

\multirow{2}*{\texttt{GPT-3.5}}
& \multirow{2}*{P0}
& \multirow{2}*{\cmark}
& $1$
& $9.89$\texttt{\%}
& $18$ / $182$
& $32.22$\texttt{\%}
& $298$ / $925$
\\

&&& $5$
& $8.79$\texttt{\%}
& $16$ / $182$
& $41.30$\texttt{\%}
& $382$ / $925$
\\

\midrule

\multirow{2}*{\texttt{GPT-3.5}}
& \multirow{2}*{P4}
& \multirow{2}*{\xmark}
& $1$
& $9.89$\texttt{\%}
& $18$ / $182$
& $4.97$\texttt{\%}
& $46$ / $925$
\\

&&& $5$
& $32.42$\texttt{\%}
& $59$ / $182$
& $13.95$\texttt{\%}
& $129$ / $925$
\\

\bottomrule
\end{tabular}
\vspace{0.5em}
\caption{Ablations on the top-100 Rust packages.}
\label{tab:result_rustcrate_ablation}
\vspace{-2em}
\end{table}

%% file: tables/prompt_variants.tex
\begin{table}[t]
\centering
\setlength{\tabcolsep}{2.2pt}
\begin{tabular}{cl cc ccc}
& & & & \multicolumn{3}{c}{\texttt{\#}\textbf{Failures}}\\
\cmidrule(lr){5-7}

\multicolumn{2}{c}{\textbf{Prompt Variant}}
& \textbf{Fix}\texttt{\%}
& \texttt{\#}\textbf{Fixed}
& \textbf{Format}
& \textbf{Build}
& \textbf{Test}\\
\toprule

P1
& \texttt{ChangeLog-basic}
& $10.74$\texttt{\%}
& $29$ / $270$
& $236$
& $2$
& $3$
\\

P2
& \texttt{Line prefixes}
& $24.07$\texttt{\%}
& $65$ / $270$
& $197$
& $4$
& $4$
\\

P3
& \texttt{Localization}
& $58.15$\texttt{\%}
& $157$ / $270$
& $67$
& $35$
& $11$
\\

P4
& \texttt{Description first}
& $73.70$\texttt{\%}
& $199$ / $270$
& $44$
& $10$
& $17$
\\

\bottomrule
\end{tabular}
\vspace{0.5em}
\caption{Evaluating variants of the \tool prompt template on the Rust error code
micro-benchmarks with GPT-3.5.}
\label{tab:pvariants}
\vspace{-2em}
\end{table}

%% file: tables/linter.tex
\begin{table}[t]
\centering
\begin{tabular}{r cccc}
& \multicolumn{2}{c}{\textbf{Clippy}}
& \multicolumn{2}{c}{\textbf{\tool}}\\
\cmidrule(lr){2-3}\cmidrule(lr){4-5}

\textbf{Category}
& \textbf{Fix}\texttt{\%}
& \texttt{\#}\textbf{Fixed}
& \textbf{Fix}\texttt{\%}
& \texttt{\#}\textbf{Fixed}\\
\toprule

Complexity
& $50.00$\texttt{\%}
& $17$ / $34$
& $\textbf{91.18}$\texttt{\%}
& $\textbf{31}$ \textbf{/} $\textbf{34}$
\\

Pedantic
& $26.12$\texttt{\%}
& $76$ / $291$
& $\textbf{71.48}$\texttt{\%}
& $\textbf{208}$ \textbf{/} $\textbf{291}$
\\

Style
& $76.19$\texttt{\%}
& $16$ / $21$
& $\textbf{95.24}$\texttt{\%}
& $\textbf{20}$ \textbf{/} $\textbf{21}$
\\

\midrule

Total
& $31.50$\texttt{\%}
& $109$ / $346$
& $\textbf{74.86}$\texttt{\%}
& $\textbf{259}$ \textbf{/} $\textbf{346}$
\\

\bottomrule
\end{tabular}
\vspace{0.5em}
\caption{Evaluating \tool (GPT-4) against Clippy's auto-fix on the top-10 Rust packages.}
\label{tab:result_linter}
\vspace{-2em}
\end{table}

%% file: tables/manualevaluation.tex
\begin{table}[t]
\centering
\setlength{\tabcolsep}{4pt}
\begin{tabular}{l c}
& \textbf{\# Commits} \\
\toprule
\textbf{Unambiguous} & $55$ \\
\textbf{Matching} & $41$ \\
\textbf{Non-matching, same runtime behaviour} & $29$ \\
\textbf{Different runtime behavior} & $9$ \\
\bottomrule
\end{tabular}
\caption{Analysis of fixed commits in the top-100 benchmark}
\label{tab:manual_evaluation}
\end{table}

%% file: threats.tex
\section{Threats to Validity}
\label{Se:Threats}

\paragraph{Internal Threats} One internal threat to validity arises from the
qualitative examination of the fixes generated by \tool\ to assess their
semantic correctness (RQ4, Section~\ref{sec:evaluation}). To address this
concern, we implemented a structured
consensus-based manual evaluation involving multiple evaluators, ensuring more
reliable and consistent assessments. Another potential internal threat is data
contamination, where it might be possible that fixes to the compilation issues
that we mined from open source might have already been included in the training
data of the LLMs that we used. There is no ideal way to completely remove
contamination without sacrificing real-world scenarios, given the scope of
training data that is consumed for these models today. However, the fixes,
especially for the top-100 benchmarks were never presented online in the form of
a fix or alongside the corresponding compiler error, to the best of our
knowledge. Only the fixed version of the code might appear in a later version of
the repository.  

\begin{sloppypar}
Additionally, the use of handwritten test cases for
micro-benchmarks and Stack Overflow code snippets to verify the generated fixes
introduces a potential threat to construct validity. To counter this, we designed
the testcases
independently from the \tool's implementation.
\end{sloppypar}

Our evaluation on Rust Clippy errors, while it demonstrates the
generalizability of \tool to non-compiler errors, was limited to checking if
the resulting fixes passed clippy. A qualitative assessment of the generated fix,
say, by checking if existing tests continue to pass, or via manual inspection, would be
necessary to claim that \tool can be readily adopted for fixing clippy errors.

\paragraph{External Threats} While our evaluation of \tool\ encompasses an
extensive analysis of three Rust error datasets from popular sources, we
acknowledge that the generalizability of our findings to different datasets may
vary.

Furthermore, the nature of API access from OpenAI implies that the LLM
performance can vary over time as the models may get updated without any prior
intimation, which can impact \tool's fix accuracy.

%% file: related.tex
\section{Related Work}
\label{Se:RelatedWork}

The area of Automated Program Repair (APR) is concerned with the
problem of taking as input a ``buggy'' code fragment as well as a correctness
specification, and producing correct code as output. This process is very
fundamental to software engineering and, consequently, has received much
attention from the research community
\cite{DBLP:journals/cacm/GouesPR19,DBLP:journals/csur/Monperrus18,DBLP:journals/corr/abs-2303-18184}. Our
work can be considered as an instance of the APR problem, where the buggy code
is the current version of Rust code on a developer's machine that does not compile, 
and the correctness specification is to pass the compiler. We elaborate on a
comparison between our approach and existing APR solutions. 

In terms of techniques, some APR solutions are based on classical (non-learning-based)
techniques. Examples include search-based techniques 
\cite{DBLP:conf/icse/WenCWHC18,DBLP:conf/issta/JiangXZGC18,DBLP:journals/tosem/MechtaevGTR18,DBLP:journals/tse/YuanB20,DBLP:conf/wcre/LiuK0B19,DBLP:conf/icse/TanR15,DBLP:conf/icse/TanDGR18,transplanfix} 
that have a pre-defined space of
potential \textit{patches} and must find one that works in this space.
Semantics-based techniques
\cite{DBLP:conf/icse/NguyenQRC13,DBLP:conf/icse/MechtaevYR16,DBLP:journals/tse/XuanMDCMDBM17,Liu_2020}
formulate the repair as a constraint that must be
solved. Symbolic techniques rely on manually-designed transformations for
constructing the fix 
\cite{DBLP:conf/icse/TonderG18,DBLP:conf/sp/0002LTJ19,DBLP:journals/corr/abs-2108-02490}.
These techniques, by construction, are limited in the space of possible repairs that they
consider. Learning-based approaches
\cite{DBLP:conf/kbse/WhiteTVP16,DBLP:conf/aaai/GuptaPKS17,DBLP:conf/icse/Li0N20,DBLP:conf/sigsoft/ZhuSXZY0Z21,DBLP:conf/icse/JiangL021,DBLP:conf/icse/YeMM22,DBLP:journals/tosem/TufanoWBPWP19,katana,9825852,knod,SynShine} 
overcome this limitation by leveraging deep learning, however, they require
supervised training data (pairs of buggy and patched code) which is
time-consuming and expensive to set up. $\tool$, on the other hand, does not
need training, or even fine-tuning, thus skipping data collection altogether. 
It instead uses the latest pre-trained LLMs; these models are
trained on a massive scale, and have the ability of following generic
instructions \cite{DBLP:conf/nips/Ouyang0JAWMZASR22}.

The potential of LLMs as powerful APR agents has been acknowledged in previous
studies
\cite{DBLP:journals/corr/abs-2208-11640,DBLP:journals/corr/abs-2303-07263,DBLP:conf/sp/PearceTAKD23,DBLP:journals/corr/abs-2112-02125,DBLP:conf/icse-apr/PrennerBR22,DBLP:conf/icse/XiaWZ23,DBLP:conf/icse/FanGMRT23,xia2022practical}.
For instance, Xia et al. \cite{xia2022practical} conducts an extensive study on
the application of nine state-of-the-art pre-trained language models (PLMs) for
APR on datasets from three different languages. They demonstrate that PLMs
outperform existing APR techniques, with larger models achieving better
performance. Fan et al. \cite{DBLP:conf/icse/FanGMRT23} evaluates the Codex edit
model \cite{codexe} on a Java defects dataset from LeetCode \cite{leetcode}.
Prenner et al. \cite{DBLP:conf/icse-apr/PrennerBR22} explores Java and Python
implementations of buggy algorithms, generating complete fixed functions using
LLMs. Building on this line of work, Xia et al. \cite{DBLP:conf/icse/XiaWZ23}
experiments with newer LLMs (GPT-3 series, CodeT5
\cite{DBLP:conf/emnlp/0034WJH21}, and InCoder
\cite{DBLP:conf/iclr/FriedAL0WSZYZL23}), a larger set of benchmarks (such as
Defects4J \cite{DBLP:conf/issta/JustJE14}), and compares against multiple APR
tools. Additionally, researchers have investigated the potential of APR
techniques to enhance the reliability of code generated by LLMs. Jesse et al.
\cite{10174227} examines the generation of single statement bugs (SStuBs) by
Codex and propose avoidance strategies, while Fan et al. \cite{10172854}
systematically studies the use of automated program repair techniques to fix
incorrect solutions produced by LLMs in LeetCode contests.

The above techniques differ from our work in
multiple dimensions. First, they rely exclusive on the model to produce the
patch, whereas we use a pipeline that iterates with the compiler to arrive at the
fix. Second, our focus on Rust is unique. There is relatively much lesser code in Rust 
compared to Java and Python that the above work had used. It is not immediately evident 
if LLM-based techniques will carry over to Rust without impacting their accuracy,
justifying the need to study Rust errors. Third, we consider 
compiler errors as opposed to above work that considered multiple
kinds of functional errors with the requirement of passing given test cases.
These are different kinds of errors, and moreover, our work does not require the
presence of a test suite, making it readily deployable in any scenario where the
user is ready to build their code. Furthermore, we leverage the specific nature
of errors, namely, compiler-generated error messages, for crafting the prompt and improving accuracy. 

There is also work on fixing statically-detected errors, as opposed to fixing of failing test cases, 
For instance, RING
\cite{DBLP:journals/corr/abs-2208-11640} considers retrieval-augmented few-shot
prompting to fix syntactic errors in multiple languages. It divides the bug-fixing process into three stages: fault localization using language tooling, program transformation through few-shot learning with LLMs, and candidate ranking based on token probabilities. RING effectively leverages developer intuition and LLM capabilities to address various types of bugs without requiring user intervention.
InferFix \cite{DBLP:journals/corr/abs-2303-07263} uses an LLM to fix errors reported
by a static analysis tool (CodeQL). They rely on fine-tuning Codex for improved
accuracy. Pearce et al. \cite{DBLP:conf/sp/PearceTAKD23} performs a large scale study to explore the potential of LLMs in automatically repairing cybersecurity bugs. They investigate the use of zero-shot vulnerability repair with five commercially available LLMs, an open-source model, and a locally-trained model, which are evaluated on a mix of synthetic, handcrafted, and real-world security bug scenarios.
FitRepair \cite{fitrepair} combines LLMs in cloze-style APR with insights from the plastic surgery hypothesis. Their method involves training two separate models using innovative fine-tuning strategies: Knowledge-Intensified fine-tuning, and Repair-Oriented fine-tuning. Additionally, it introduces a Relevant-Identifier prompting strategy by using information retrieval and static analysis to obtain a list of relevant/rare identifiers not seen in the model's immediate context. Our work does not rely on
fine-tuning, and instead utilizes iterative fixing with instruction-tuned LLMs.
We do, however, believe that \tool can benefit from few-shot prompting where
similar examples of fixing a particular compiler error as provided as part of
the prompt \cite{DBLP:conf/icse/NashidSM23}. We leave this direction as future work.

%% file: conclusions.tex
\section{Conclusions}
\label{sec:conclusions}

This paper presents \tool as a tool for automatically generating patches for
compilation errors in Rust. \tool leverages emergent capabilities of Pre-Trained Large
Language Models to deliver impressive results. It demonstrates that the latest
advancements in LLMs (e.g., GPT-4 over GPT-3.5) as well as combining them with formal
tools such as a compiler leads to a very effective solution for fixing code
errors. 

LLMs are sensitive to the prompts that they are supplied. We demonstrate
the features that were needed to help the model communicate code
changes, bring accuracy up from a mere $10\%$ to nearly $74\%$. We further
demonstrate generality of \tool by auto-fixing rust lint errors. 

This evidence should add encouragement to the wave of building LLM-powered tools
for software engineering. We plan to release our dataset to enable further research.

%% file: paper.bbl
%%% -*-BibTeX-*-
%%% Do NOT edit. File created by BibTeX with style
%%% ACM-Reference-Format-Journals [18-Jan-2012].

\begin{thebibliography}{71}

%%% ====================================================================
%%% NOTE TO THE USER: you can override these defaults by providing
%%% customized versions of any of these macros before the \bibliography
%%% command.  Each of them MUST provide its own final punctuation,
%%% except for \shownote{}, \showDOI{}, and \showURL{}.  The latter two
%%% do not use final punctuation, in order to avoid confusing it with
%%% the Web address.
%%%
%%% To suppress output of a particular field, define its macro to expand
%%% to an empty string, or better, \unskip, like this:
%%%
%%% \newcommand{\showDOI}[1]{\unskip}   % LaTeX syntax
%%%
%%% \def \showDOI #1{\unskip}           % plain TeX syntax
%%%
%%% ====================================================================

\ifx \showCODEN    \undefined \def \showCODEN     #1{\unskip}     \fi
\ifx \showDOI      \undefined \def \showDOI       #1{#1}\fi
\ifx \showISBNx    \undefined \def \showISBNx     #1{\unskip}     \fi
\ifx \showISBNxiii \undefined \def \showISBNxiii  #1{\unskip}     \fi
\ifx \showISSN     \undefined \def \showISSN      #1{\unskip}     \fi
\ifx \showLCCN     \undefined \def \showLCCN      #1{\unskip}     \fi
\ifx \shownote     \undefined \def \shownote      #1{#1}          \fi
\ifx \showarticletitle \undefined \def \showarticletitle #1{#1}   \fi
\ifx \showURL      \undefined \def \showURL       {\relax}        \fi
% The following commands are used for tagged output and should be
% invisible to TeX
\providecommand\bibfield[2]{#2}
\providecommand\bibinfo[2]{#2}
\providecommand\natexlab[1]{#1}
\providecommand\showeprint[2][]{arXiv:#2}

\bibitem[Ahmed et~al\mbox{.}(2023)]%
        {SynShine}
\bibfield{author}{\bibinfo{person}{Toufique Ahmed}, \bibinfo{person}{Noah~Rose
  Ledesma}, {and} \bibinfo{person}{Premkumar Devanbu}.}
  \bibinfo{year}{2023}\natexlab{}.
\newblock \showarticletitle{SynShine: Improved Fixing of Syntax Errors}.
\newblock \bibinfo{journal}{\emph{IEEE Transactions on Software Engineering}}
  \bibinfo{volume}{49}, \bibinfo{number}{4} (\bibinfo{year}{2023}),
  \bibinfo{pages}{2169--2181}.
\newblock
\urldef\tempurl%
\url{https://doi.org/10.1109/TSE.2022.3212635}
\showDOI{\tempurl}


\bibitem[Anil et~al\mbox{.}(2023)]%
        {DBLP:journals/corr/abs-2305-10403}
\bibfield{author}{\bibinfo{person}{Rohan Anil}, \bibinfo{person}{Andrew~M.
  Dai}, \bibinfo{person}{Orhan Firat}, \bibinfo{person}{Melvin Johnson},
  \bibinfo{person}{Dmitry Lepikhin}, \bibinfo{person}{Alexandre Passos}, {and}
  \bibinfo{person}{et al.}} \bibinfo{year}{2023}\natexlab{}.
\newblock \showarticletitle{PaLM 2 Technical Report}.
\newblock \bibinfo{journal}{\emph{CoRR}}  \bibinfo{volume}{abs/2305.10403}
  (\bibinfo{year}{2023}).
\newblock
\urldef\tempurl%
\url{https://doi.org/10.48550/arXiv.2305.10403}
\showDOI{\tempurl}
\showeprint[arXiv]{2305.10403}


\bibitem[{AWS}(2022)]%
        {amazonrust}
\bibfield{author}{\bibinfo{person}{{AWS}}.} \bibinfo{year}{2022}\natexlab{}.
\newblock \bibinfo{title}{Sustainability with Rust}.
\newblock
  \bibinfo{howpublished}{\url{https://aws.amazon.com/blogs/opensource/sustainability-with-rust/}}.
\newblock


\bibitem[Connor et~al\mbox{.}(2022)]%
        {9825852}
\bibfield{author}{\bibinfo{person}{Aidan Connor}, \bibinfo{person}{Aaron
  Harris}, \bibinfo{person}{Nathan Cooper}, {and} \bibinfo{person}{Denys
  Poshyvanyk}.} \bibinfo{year}{2022}\natexlab{}.
\newblock \showarticletitle{Can We Automatically Fix Bugs by Learning Edit
  Operations?}. In \bibinfo{booktitle}{\emph{2022 IEEE International Conference
  on Software Analysis, Evolution and Reengineering (SANER)}}.
  \bibinfo{pages}{782--792}.
\newblock
\urldef\tempurl%
\url{https://doi.org/10.1109/SANER53432.2022.00096}
\showDOI{\tempurl}


\bibitem[Costea et~al\mbox{.}(2021)]%
        {DBLP:journals/corr/abs-2108-02490}
\bibfield{author}{\bibinfo{person}{Andreea Costea}, \bibinfo{person}{Abhishek
  Tiwari}, \bibinfo{person}{Sigmund Chianasta}, \bibinfo{person}{Kishore R},
  \bibinfo{person}{Abhik Roychoudhury}, {and} \bibinfo{person}{Ilya Sergey}.}
  \bibinfo{year}{2021}\natexlab{}.
\newblock \showarticletitle{{HIPPODROME:} Data Race Repair using Static
  Analysis Summaries}.
\newblock \bibinfo{journal}{\emph{CoRR}}  \bibinfo{volume}{abs/2108.02490}
  (\bibinfo{year}{2021}).
\newblock
\showeprint[arXiv]{2108.02490}
\urldef\tempurl%
\url{https://arxiv.org/abs/2108.02490}
\showURL{%
\tempurl}


\bibitem[{Emery Berger}(2023)]%
        {chatdbg}
\bibfield{author}{\bibinfo{person}{{Emery Berger}}.}
  \bibinfo{year}{2023}\natexlab{}.
\newblock \bibinfo{title}{ChatDBG}.
\newblock
  \bibinfo{howpublished}{\url{https://github.com/plasma-umass/ChatDBG}}.
\newblock


\bibitem[Fan et~al\mbox{.}(2023a)]%
        {DBLP:conf/icse/FanGMRT23}
\bibfield{author}{\bibinfo{person}{Zhiyu Fan}, \bibinfo{person}{Xiang Gao},
  \bibinfo{person}{Martin Mirchev}, \bibinfo{person}{Abhik Roychoudhury}, {and}
  \bibinfo{person}{Shin~Hwei Tan}.} \bibinfo{year}{2023}\natexlab{a}.
\newblock \showarticletitle{Automated Repair of Programs from Large Language
  Models}. In \bibinfo{booktitle}{\emph{45th {IEEE/ACM} International
  Conference on Software Engineering, {ICSE} 2023, Melbourne, Australia, May
  14-20, 2023}}. \bibinfo{publisher}{{IEEE}}, \bibinfo{pages}{1469--1481}.
\newblock
\urldef\tempurl%
\url{https://doi.org/10.1109/ICSE48619.2023.00128}
\showDOI{\tempurl}


\bibitem[Fan et~al\mbox{.}(2023b)]%
        {10172854}
\bibfield{author}{\bibinfo{person}{Zhiyu Fan}, \bibinfo{person}{Xiang Gao},
  \bibinfo{person}{Martin Mirchev}, \bibinfo{person}{Abhik Roychoudhury}, {and}
  \bibinfo{person}{Shin~Hwei Tan}.} \bibinfo{year}{2023}\natexlab{b}.
\newblock \showarticletitle{Automated Repair of Programs from Large Language
  Models}. In \bibinfo{booktitle}{\emph{2023 IEEE/ACM 45th International
  Conference on Software Engineering (ICSE)}}. \bibinfo{pages}{1469--1481}.
\newblock
\urldef\tempurl%
\url{https://doi.org/10.1109/ICSE48619.2023.00128}
\showDOI{\tempurl}


\bibitem[Fried et~al\mbox{.}(2023)]%
        {DBLP:conf/iclr/FriedAL0WSZYZL23}
\bibfield{author}{\bibinfo{person}{Daniel Fried}, \bibinfo{person}{Armen
  Aghajanyan}, \bibinfo{person}{Jessy Lin}, \bibinfo{person}{Sida Wang},
  \bibinfo{person}{Eric Wallace}, \bibinfo{person}{Freda Shi},
  \bibinfo{person}{Ruiqi Zhong}, \bibinfo{person}{Scott Yih},
  \bibinfo{person}{Luke Zettlemoyer}, {and} \bibinfo{person}{Mike Lewis}.}
  \bibinfo{year}{2023}\natexlab{}.
\newblock \showarticletitle{InCoder: {A} Generative Model for Code Infilling
  and Synthesis}. In \bibinfo{booktitle}{\emph{The Eleventh International
  Conference on Learning Representations, {ICLR} 2023, Kigali, Rwanda, May 1-5,
  2023}}. \bibinfo{publisher}{OpenReview.net}.
\newblock
\urldef\tempurl%
\url{https://openreview.net/pdf?id=hQwb-lbM6EL}
\showURL{%
\tempurl}


\bibitem[{GitHub}(2022)]%
        {copilot}
\bibfield{author}{\bibinfo{person}{{GitHub}}.} \bibinfo{year}{2022}\natexlab{}.
\newblock \bibinfo{title}{GitHub Copilot}.
\newblock \bibinfo{howpublished}{\url{https://github.com/features/copilot}}.
\newblock


\bibitem[{GitHub}(2023)]%
        {copilotx}
\bibfield{author}{\bibinfo{person}{{GitHub}}.} \bibinfo{year}{2023}\natexlab{}.
\newblock \bibinfo{title}{GitHub Copilot-X}.
\newblock
  \bibinfo{howpublished}{\url{https://github.com/features/preview/copilot-x}}.
\newblock


\bibitem[Goues et~al\mbox{.}(2019)]%
        {DBLP:journals/cacm/GouesPR19}
\bibfield{author}{\bibinfo{person}{Claire~Le Goues}, \bibinfo{person}{Michael
  Pradel}, {and} \bibinfo{person}{Abhik Roychoudhury}.}
  \bibinfo{year}{2019}\natexlab{}.
\newblock \showarticletitle{Automated program repair}.
\newblock \bibinfo{journal}{\emph{Commun. {ACM}}} \bibinfo{volume}{62},
  \bibinfo{number}{12} (\bibinfo{year}{2019}), \bibinfo{pages}{56--65}.
\newblock
\urldef\tempurl%
\url{https://doi.org/10.1145/3318162}
\showDOI{\tempurl}


\bibitem[Gupta et~al\mbox{.}(2017)]%
        {DBLP:conf/aaai/GuptaPKS17}
\bibfield{author}{\bibinfo{person}{Rahul Gupta}, \bibinfo{person}{Soham Pal},
  \bibinfo{person}{Aditya Kanade}, {and} \bibinfo{person}{Shirish~K. Shevade}.}
  \bibinfo{year}{2017}\natexlab{}.
\newblock \showarticletitle{DeepFix: Fixing Common {C} Language Errors by Deep
  Learning}. In \bibinfo{booktitle}{\emph{Proceedings of the Thirty-First
  {AAAI} Conference on Artificial Intelligence, February 4-9, 2017, San
  Francisco, California, {USA}}}, \bibfield{editor}{\bibinfo{person}{Satinder
  Singh} {and} \bibinfo{person}{Shaul Markovitch}} (Eds.).
  \bibinfo{publisher}{{AAAI} Press}, \bibinfo{pages}{1345--1351}.
\newblock
\urldef\tempurl%
\url{http://aaai.org/ocs/index.php/AAAI/AAAI17/paper/view/14603}
\showURL{%
\tempurl}


\bibitem[Huang et~al\mbox{.}(2023)]%
        {DBLP:journals/corr/abs-2303-18184}
\bibfield{author}{\bibinfo{person}{Kai Huang}, \bibinfo{person}{Zhengzi Xu},
  \bibinfo{person}{Su Yang}, \bibinfo{person}{Hongyu Sun},
  \bibinfo{person}{Xuejun Li}, \bibinfo{person}{Zheng Yan}, {and}
  \bibinfo{person}{Yuqing Zhang}.} \bibinfo{year}{2023}\natexlab{}.
\newblock \showarticletitle{A Survey on Automated Program Repair Techniques}.
\newblock \bibinfo{journal}{\emph{CoRR}}  \bibinfo{volume}{abs/2303.18184}
  (\bibinfo{year}{2023}).
\newblock
\urldef\tempurl%
\url{https://doi.org/10.48550/arXiv.2303.18184}
\showDOI{\tempurl}
\showeprint[arXiv]{2303.18184}


\bibitem[Huang et~al\mbox{.}(2019)]%
        {DBLP:conf/sp/0002LTJ19}
\bibfield{author}{\bibinfo{person}{Zhen Huang}, \bibinfo{person}{David Lie},
  \bibinfo{person}{Gang Tan}, {and} \bibinfo{person}{Trent Jaeger}.}
  \bibinfo{year}{2019}\natexlab{}.
\newblock \showarticletitle{Using Safety Properties to Generate Vulnerability
  Patches}. In \bibinfo{booktitle}{\emph{2019 {IEEE} Symposium on Security and
  Privacy, {SP} 2019, San Francisco, CA, USA, May 19-23, 2019}}.
  \bibinfo{publisher}{{IEEE}}, \bibinfo{pages}{539--554}.
\newblock
\urldef\tempurl%
\url{https://doi.org/10.1109/SP.2019.00071}
\showDOI{\tempurl}


\bibitem[jeromefroe(2016)]%
        {rust-so-question}
\bibfield{author}{\bibinfo{person}{jeromefroe}.}
  \bibinfo{year}{2016}\natexlab{}.
\newblock \bibinfo{title}{Question about a {Rust} compilation error on {Stack
  Overflow}}.
\newblock
  \bibinfo{howpublished}{\url{https://stackoverflow.com/questions/40299671}}.
\newblock


\bibitem[Jesse et~al\mbox{.}(2023)]%
        {10174227}
\bibfield{author}{\bibinfo{person}{Kevin Jesse}, \bibinfo{person}{Toufique
  Ahmed}, \bibinfo{person}{Premkumar~T. Devanbu}, {and} \bibinfo{person}{Emily
  Morgan}.} \bibinfo{year}{2023}\natexlab{}.
\newblock \showarticletitle{Large Language Models and Simple, Stupid Bugs}. In
  \bibinfo{booktitle}{\emph{2023 IEEE/ACM 20th International Conference on
  Mining Software Repositories (MSR)}}. \bibinfo{pages}{563--575}.
\newblock
\urldef\tempurl%
\url{https://doi.org/10.1109/MSR59073.2023.00082}
\showDOI{\tempurl}


\bibitem[Jiang et~al\mbox{.}(2018)]%
        {DBLP:conf/issta/JiangXZGC18}
\bibfield{author}{\bibinfo{person}{Jiajun Jiang}, \bibinfo{person}{Yingfei
  Xiong}, \bibinfo{person}{Hongyu Zhang}, \bibinfo{person}{Qing Gao}, {and}
  \bibinfo{person}{Xiangqun Chen}.} \bibinfo{year}{2018}\natexlab{}.
\newblock \showarticletitle{Shaping program repair space with existing patches
  and similar code}. In \bibinfo{booktitle}{\emph{Proceedings of the 27th {ACM}
  {SIGSOFT} International Symposium on Software Testing and Analysis, {ISSTA}
  2018, Amsterdam, The Netherlands, July 16-21, 2018}},
  \bibfield{editor}{\bibinfo{person}{Frank Tip} {and} \bibinfo{person}{Eric
  Bodden}} (Eds.). \bibinfo{publisher}{{ACM}}, \bibinfo{pages}{298--309}.
\newblock
\urldef\tempurl%
\url{https://doi.org/10.1145/3213846.3213871}
\showDOI{\tempurl}


\bibitem[Jiang et~al\mbox{.}(2023)]%
        {knod}
\bibfield{author}{\bibinfo{person}{Nan Jiang}, \bibinfo{person}{Thibaud
  Lutellier}, \bibinfo{person}{Yiling Lou}, \bibinfo{person}{Lin Tan},
  \bibinfo{person}{Dan Goldwasser}, {and} \bibinfo{person}{Xiangyu Zhang}.}
  \bibinfo{year}{2023}\natexlab{}.
\newblock \bibinfo{title}{KNOD: Domain Knowledge Distilled Tree Decoder for
  Automated Program Repair}.
\newblock
\newblock
\showeprint[arxiv]{2302.01857}~[cs.SE]


\bibitem[Jiang et~al\mbox{.}(2021)]%
        {DBLP:conf/icse/JiangL021}
\bibfield{author}{\bibinfo{person}{Nan Jiang}, \bibinfo{person}{Thibaud
  Lutellier}, {and} \bibinfo{person}{Lin Tan}.}
  \bibinfo{year}{2021}\natexlab{}.
\newblock \showarticletitle{{CURE:} Code-Aware Neural Machine Translation for
  Automatic Program Repair}. In \bibinfo{booktitle}{\emph{43rd {IEEE/ACM}
  International Conference on Software Engineering, {ICSE} 2021, Madrid, Spain,
  22-30 May 2021}}. \bibinfo{publisher}{{IEEE}}, \bibinfo{pages}{1161--1173}.
\newblock
\urldef\tempurl%
\url{https://doi.org/10.1109/ICSE43902.2021.00107}
\showDOI{\tempurl}


\bibitem[Jin et~al\mbox{.}(2023)]%
        {DBLP:journals/corr/abs-2303-07263}
\bibfield{author}{\bibinfo{person}{Matthew Jin}, \bibinfo{person}{Syed
  Shahriar}, \bibinfo{person}{Michele Tufano}, \bibinfo{person}{Xin Shi},
  \bibinfo{person}{Shuai Lu}, \bibinfo{person}{Neel Sundaresan}, {and}
  \bibinfo{person}{Alexey Svyatkovskiy}.} \bibinfo{year}{2023}\natexlab{}.
\newblock \showarticletitle{InferFix: End-to-End Program Repair with LLMs}.
\newblock \bibinfo{journal}{\emph{CoRR}}  \bibinfo{volume}{abs/2303.07263}
  (\bibinfo{year}{2023}).
\newblock
\urldef\tempurl%
\url{https://doi.org/10.48550/arXiv.2303.07263}
\showDOI{\tempurl}
\showeprint[arXiv]{2303.07263}


\bibitem[Joshi et~al\mbox{.}(2022)]%
        {DBLP:journals/corr/abs-2208-11640}
\bibfield{author}{\bibinfo{person}{Harshit Joshi}, \bibinfo{person}{Jos{\'{e}}
  Pablo~Cambronero S{\'{a}}nchez}, \bibinfo{person}{Sumit Gulwani},
  \bibinfo{person}{Vu Le}, \bibinfo{person}{Ivan Radicek}, {and}
  \bibinfo{person}{Gust Verbruggen}.} \bibinfo{year}{2022}\natexlab{}.
\newblock \showarticletitle{Repair Is Nearly Generation: Multilingual Program
  Repair with LLMs}.
\newblock \bibinfo{journal}{\emph{CoRR}}  \bibinfo{volume}{abs/2208.11640}
  (\bibinfo{year}{2022}).
\newblock
\urldef\tempurl%
\url{https://doi.org/10.48550/arXiv.2208.11640}
\showDOI{\tempurl}
\showeprint[arXiv]{2208.11640}


\bibitem[Just et~al\mbox{.}(2014)]%
        {DBLP:conf/issta/JustJE14}
\bibfield{author}{\bibinfo{person}{Ren{\'{e}} Just}, \bibinfo{person}{Darioush
  Jalali}, {and} \bibinfo{person}{Michael~D. Ernst}.}
  \bibinfo{year}{2014}\natexlab{}.
\newblock \showarticletitle{Defects4J: a database of existing faults to enable
  controlled testing studies for Java programs}. In
  \bibinfo{booktitle}{\emph{International Symposium on Software Testing and
  Analysis, {ISSTA} '14, San Jose, CA, {USA} - July 21 - 26, 2014}},
  \bibfield{editor}{\bibinfo{person}{Corina~S. Pasareanu} {and}
  \bibinfo{person}{Darko Marinov}} (Eds.). \bibinfo{publisher}{{ACM}},
  \bibinfo{pages}{437--440}.
\newblock
\urldef\tempurl%
\url{https://doi.org/10.1145/2610384.2628055}
\showDOI{\tempurl}


\bibitem[{LeetCode}(2023)]%
        {leetcode}
\bibfield{author}{\bibinfo{person}{{LeetCode}}.}
  \bibinfo{year}{2023}\natexlab{}.
\newblock \bibinfo{title}{LeetCode Contest}.
\newblock \bibinfo{howpublished}{\url{https://leetcode.com/contest}}.
\newblock


\bibitem[Li et~al\mbox{.}(2020)]%
        {DBLP:conf/icse/Li0N20}
\bibfield{author}{\bibinfo{person}{Yi Li}, \bibinfo{person}{Shaohua Wang},
  {and} \bibinfo{person}{Tien~N. Nguyen}.} \bibinfo{year}{2020}\natexlab{}.
\newblock \showarticletitle{DLFix: context-based code transformation learning
  for automated program repair}. In \bibinfo{booktitle}{\emph{{ICSE} '20: 42nd
  International Conference on Software Engineering, Seoul, South Korea, 27 June
  - 19 July, 2020}}, \bibfield{editor}{\bibinfo{person}{Gregg Rothermel} {and}
  \bibinfo{person}{Doo{-}Hwan Bae}} (Eds.). \bibinfo{publisher}{{ACM}},
  \bibinfo{pages}{602--614}.
\newblock
\urldef\tempurl%
\url{https://doi.org/10.1145/3377811.3380345}
\showDOI{\tempurl}


\bibitem[{Linux kernel development community}(2020)]%
        {rustlinux}
\bibfield{author}{\bibinfo{person}{{Linux kernel development community}}.}
  \bibinfo{year}{2020}\natexlab{}.
\newblock \bibinfo{title}{Rust in Linux Kernel}.
\newblock
  \bibinfo{howpublished}{\url{https://docs.kernel.org/next/rust/index.html}}.
\newblock


\bibitem[Liu et~al\mbox{.}(2019)]%
        {DBLP:conf/wcre/LiuK0B19}
\bibfield{author}{\bibinfo{person}{Kui Liu}, \bibinfo{person}{Anil Koyuncu},
  \bibinfo{person}{Dongsun Kim}, {and} \bibinfo{person}{Tegawend{\'{e}}~F.
  Bissyand{\'{e}}}.} \bibinfo{year}{2019}\natexlab{}.
\newblock \showarticletitle{{AVATAR:} Fixing Semantic Bugs with Fix Patterns of
  Static Analysis Violations}. In \bibinfo{booktitle}{\emph{26th {IEEE}
  International Conference on Software Analysis, Evolution and Reengineering,
  {SANER} 2019, Hangzhou, China, February 24-27, 2019}},
  \bibfield{editor}{\bibinfo{person}{Xinyu Wang}, \bibinfo{person}{David Lo},
  {and} \bibinfo{person}{Emad Shihab}} (Eds.). \bibinfo{publisher}{{IEEE}},
  \bibinfo{pages}{456--467}.
\newblock
\urldef\tempurl%
\url{https://doi.org/10.1109/SANER.2019.8667970}
\showDOI{\tempurl}


\bibitem[Liu et~al\mbox{.}(2020)]%
        {Liu_2020}
\bibfield{author}{\bibinfo{person}{Kui Liu}, \bibinfo{person}{Shangwen Wang},
  \bibinfo{person}{Anil Koyuncu}, \bibinfo{person}{Kisub Kim},
  \bibinfo{person}{Tegawend{\'{e}}~F. Bissyand{\'{e}}},
  \bibinfo{person}{Dongsun Kim}, \bibinfo{person}{Peng Wu},
  \bibinfo{person}{Jacques Klein}, \bibinfo{person}{Xiaoguang Mao}, {and}
  \bibinfo{person}{Yves~Le Traon}.} \bibinfo{year}{2020}\natexlab{}.
\newblock \showarticletitle{On the efficiency of test suite based program
  repair}. In \bibinfo{booktitle}{\emph{Proceedings of the {ACM}/{IEEE} 42nd
  International Conference on Software Engineering}}. \bibinfo{publisher}{ACM}.
\newblock
\urldef\tempurl%
\url{https://doi.org/10.1145/3377811.3380338}
\showDOI{\tempurl}


\bibitem[{Mark Russinovich}(2023)]%
        {rustwindows}
\bibfield{author}{\bibinfo{person}{{Mark Russinovich}}.}
  \bibinfo{year}{2023}\natexlab{}.
\newblock \bibinfo{title}{Rust in the Windows kernel}.
\newblock
  \bibinfo{howpublished}{\url{https://twitter.com/markrussinovich/status/1656416376125538304?lang=en}}.
\newblock


\bibitem[Mechtaev et~al\mbox{.}(2018)]%
        {DBLP:journals/tosem/MechtaevGTR18}
\bibfield{author}{\bibinfo{person}{Sergey Mechtaev}, \bibinfo{person}{Xiang
  Gao}, \bibinfo{person}{Shin~Hwei Tan}, {and} \bibinfo{person}{Abhik
  Roychoudhury}.} \bibinfo{year}{2018}\natexlab{}.
\newblock \showarticletitle{Test-Equivalence Analysis for Automatic Patch
  Generation}.
\newblock \bibinfo{journal}{\emph{{ACM} Trans. Softw. Eng. Methodol.}}
  \bibinfo{volume}{27}, \bibinfo{number}{4} (\bibinfo{year}{2018}),
  \bibinfo{pages}{15:1--15:37}.
\newblock
\urldef\tempurl%
\url{https://doi.org/10.1145/3241980}
\showDOI{\tempurl}


\bibitem[Mechtaev et~al\mbox{.}(2016)]%
        {DBLP:conf/icse/MechtaevYR16}
\bibfield{author}{\bibinfo{person}{Sergey Mechtaev}, \bibinfo{person}{Jooyong
  Yi}, {and} \bibinfo{person}{Abhik Roychoudhury}.}
  \bibinfo{year}{2016}\natexlab{}.
\newblock \showarticletitle{Angelix: scalable multiline program patch synthesis
  via symbolic analysis}. In \bibinfo{booktitle}{\emph{Proceedings of the 38th
  International Conference on Software Engineering, {ICSE} 2016, Austin, TX,
  USA, May 14-22, 2016}}, \bibfield{editor}{\bibinfo{person}{Laura~K. Dillon},
  \bibinfo{person}{Willem Visser}, {and} \bibinfo{person}{Laurie~A. Williams}}
  (Eds.). \bibinfo{publisher}{{ACM}}, \bibinfo{pages}{691--701}.
\newblock
\urldef\tempurl%
\url{https://doi.org/10.1145/2884781.2884807}
\showDOI{\tempurl}


\bibitem[{Microsoft}(2023a)]%
        {bingai}
\bibfield{author}{\bibinfo{person}{{Microsoft}}.}
  \bibinfo{year}{2023}\natexlab{a}.
\newblock \bibinfo{title}{AI powered Bing}.
\newblock
  \bibinfo{howpublished}{\url{https://blogs.microsoft.com/blog/2023/02/07/reinventing-search-with-a-new-ai-powered-microsoft-bing-and-edge-your-copilot-for-the-web/}}.
\newblock


\bibitem[{Microsoft}(2023b)]%
        {officecopilot}
\bibfield{author}{\bibinfo{person}{{Microsoft}}.}
  \bibinfo{year}{2023}\natexlab{b}.
\newblock \bibinfo{title}{Microsoft 365 Copilot}.
\newblock
  \bibinfo{howpublished}{\url{https://blogs.microsoft.com/blog/2023/03/16/introducing-microsoft-365-copilot-your-copilot-for-work/}}.
\newblock


\bibitem[Monperrus(2018)]%
        {DBLP:journals/csur/Monperrus18}
\bibfield{author}{\bibinfo{person}{Martin Monperrus}.}
  \bibinfo{year}{2018}\natexlab{}.
\newblock \showarticletitle{Automatic Software Repair: {A} Bibliography}.
\newblock \bibinfo{journal}{\emph{{ACM} Comput. Surv.}} \bibinfo{volume}{51},
  \bibinfo{number}{1} (\bibinfo{year}{2018}), \bibinfo{pages}{17:1--17:24}.
\newblock
\urldef\tempurl%
\url{https://doi.org/10.1145/3105906}
\showDOI{\tempurl}


\bibitem[{MSRC Team}(2023)]%
        {mssurveymemory}
\bibfield{author}{\bibinfo{person}{{MSRC Team}}.}
  \bibinfo{year}{2023}\natexlab{}.
\newblock \bibinfo{title}{A proactive approach to more secure code}.
\newblock
  \bibinfo{howpublished}{\url{https://msrc.microsoft.com/blog/2019/07/a-proactive-approach-to-more-secure-code/}}.
\newblock


\bibitem[Nashid et~al\mbox{.}(2023)]%
        {DBLP:conf/icse/NashidSM23}
\bibfield{author}{\bibinfo{person}{Noor Nashid}, \bibinfo{person}{Mifta
  Sintaha}, {and} \bibinfo{person}{Ali Mesbah}.}
  \bibinfo{year}{2023}\natexlab{}.
\newblock \showarticletitle{Retrieval-Based Prompt Selection for Code-Related
  Few-Shot Learning}. In \bibinfo{booktitle}{\emph{45th {IEEE/ACM}
  International Conference on Software Engineering, {ICSE} 2023, Melbourne,
  Australia, May 14-20, 2023}}. \bibinfo{publisher}{{IEEE}},
  \bibinfo{pages}{2450--2462}.
\newblock
\urldef\tempurl%
\url{https://doi.org/10.1109/ICSE48619.2023.00205}
\showDOI{\tempurl}


\bibitem[Nguyen et~al\mbox{.}(2013)]%
        {DBLP:conf/icse/NguyenQRC13}
\bibfield{author}{\bibinfo{person}{Hoang Duong~Thien Nguyen},
  \bibinfo{person}{Dawei Qi}, \bibinfo{person}{Abhik Roychoudhury}, {and}
  \bibinfo{person}{Satish Chandra}.} \bibinfo{year}{2013}\natexlab{}.
\newblock \showarticletitle{SemFix: program repair via semantic analysis}. In
  \bibinfo{booktitle}{\emph{35th International Conference on Software
  Engineering, {ICSE} '13, San Francisco, CA, USA, May 18-26, 2013}},
  \bibfield{editor}{\bibinfo{person}{David Notkin}, \bibinfo{person}{Betty
  H.~C. Cheng}, {and} \bibinfo{person}{Klaus Pohl}} (Eds.).
  \bibinfo{publisher}{{IEEE} Computer Society}, \bibinfo{pages}{772--781}.
\newblock
\urldef\tempurl%
\url{https://doi.org/10.1109/ICSE.2013.6606623}
\showDOI{\tempurl}


\bibitem[Nye et~al\mbox{.}(2021)]%
        {DBLP:journals/corr/abs-2112-00114}
\bibfield{author}{\bibinfo{person}{Maxwell~I. Nye},
  \bibinfo{person}{Anders~Johan Andreassen}, \bibinfo{person}{Guy Gur{-}Ari},
  \bibinfo{person}{Henryk Michalewski}, \bibinfo{person}{Jacob Austin},
  \bibinfo{person}{David Bieber}, \bibinfo{person}{David Dohan},
  \bibinfo{person}{Aitor Lewkowycz}, \bibinfo{person}{Maarten Bosma},
  \bibinfo{person}{David Luan}, \bibinfo{person}{Charles Sutton}, {and}
  \bibinfo{person}{Augustus Odena}.} \bibinfo{year}{2021}\natexlab{}.
\newblock \showarticletitle{Show Your Work: Scratchpads for Intermediate
  Computation with Language Models}.
\newblock \bibinfo{journal}{\emph{CoRR}}  \bibinfo{volume}{abs/2112.00114}
  (\bibinfo{year}{2021}).
\newblock
\showeprint[arXiv]{2112.00114}
\urldef\tempurl%
\url{https://arxiv.org/abs/2112.00114}
\showURL{%
\tempurl}


\bibitem[{OpenAI}(2023a)]%
        {codexe}
\bibfield{author}{\bibinfo{person}{{OpenAI}}.}
  \bibinfo{year}{2023}\natexlab{a}.
\newblock \bibinfo{title}{Codex Edit Model}.
\newblock
  \bibinfo{howpublished}{\url{https://openai.com/blog/gpt-3-edit-insert}}.
\newblock


\bibitem[{OpenAI}(2023b)]%
        {gpt35}
\bibfield{author}{\bibinfo{person}{{OpenAI}}.}
  \bibinfo{year}{2023}\natexlab{b}.
\newblock \bibinfo{title}{{GPT-3.5}}.
\newblock
  \bibinfo{howpublished}{\url{https://platform.openai.com/docs/models/gpt-3-5}}.
\newblock


\bibitem[OpenAI(2023)]%
        {DBLP:journals/corr/abs-2303-08774}
\bibfield{author}{\bibinfo{person}{OpenAI}.} \bibinfo{year}{2023}\natexlab{}.
\newblock \showarticletitle{{GPT-4} Technical Report}.
\newblock \bibinfo{journal}{\emph{CoRR}}  \bibinfo{volume}{abs/2303.08774}
  (\bibinfo{year}{2023}).
\newblock
\urldef\tempurl%
\url{https://doi.org/10.48550/arXiv.2303.08774}
\showDOI{\tempurl}
\showeprint[arXiv]{2303.08774}


\bibitem[Ouyang et~al\mbox{.}(2022)]%
        {DBLP:conf/nips/Ouyang0JAWMZASR22}
\bibfield{author}{\bibinfo{person}{Long Ouyang}, \bibinfo{person}{Jeffrey Wu},
  \bibinfo{person}{Xu Jiang}, \bibinfo{person}{Diogo Almeida},
  \bibinfo{person}{Carroll~L. Wainwright}, \bibinfo{person}{Pamela Mishkin},
  {and} \bibinfo{person}{et al.}} \bibinfo{year}{2022}\natexlab{}.
\newblock \showarticletitle{Training language models to follow instructions
  with human feedback}. In \bibinfo{booktitle}{\emph{NeurIPS}}.
\newblock
\urldef\tempurl%
\url{http://papers.nips.cc/paper\_files/paper/2022/hash/b1efde53be364a73914f58805a001731-Abstract-Conference.html}
\showURL{%
\tempurl}


\bibitem[Pearce et~al\mbox{.}(2021)]%
        {DBLP:journals/corr/abs-2112-02125}
\bibfield{author}{\bibinfo{person}{Hammond Pearce}, \bibinfo{person}{Benjamin
  Tan}, \bibinfo{person}{Baleegh Ahmad}, \bibinfo{person}{Ramesh Karri}, {and}
  \bibinfo{person}{Brendan Dolan{-}Gavitt}.} \bibinfo{year}{2021}\natexlab{}.
\newblock \showarticletitle{Can OpenAI Codex and Other Large Language Models
  Help Us Fix Security Bugs?}
\newblock \bibinfo{journal}{\emph{CoRR}}  \bibinfo{volume}{abs/2112.02125}
  (\bibinfo{year}{2021}).
\newblock
\showeprint[arXiv]{2112.02125}
\urldef\tempurl%
\url{https://arxiv.org/abs/2112.02125}
\showURL{%
\tempurl}


\bibitem[Pearce et~al\mbox{.}(2023)]%
        {DBLP:conf/sp/PearceTAKD23}
\bibfield{author}{\bibinfo{person}{Hammond Pearce}, \bibinfo{person}{Benjamin
  Tan}, \bibinfo{person}{Baleegh Ahmad}, \bibinfo{person}{Ramesh Karri}, {and}
  \bibinfo{person}{Brendan Dolan{-}Gavitt}.} \bibinfo{year}{2023}\natexlab{}.
\newblock \showarticletitle{Examining Zero-Shot Vulnerability Repair with Large
  Language Models}. In \bibinfo{booktitle}{\emph{44th {IEEE} Symposium on
  Security and Privacy, {SP} 2023, San Francisco, CA, USA, May 21-25, 2023}}.
  \bibinfo{publisher}{{IEEE}}, \bibinfo{pages}{2339--2356}.
\newblock
\urldef\tempurl%
\url{https://doi.org/10.1109/SP46215.2023.10179420}
\showDOI{\tempurl}


\bibitem[Prenner et~al\mbox{.}(2022)]%
        {DBLP:conf/icse-apr/PrennerBR22}
\bibfield{author}{\bibinfo{person}{Julian~Aron Prenner}, \bibinfo{person}{Hlib
  Babii}, {and} \bibinfo{person}{Romain Robbes}.}
  \bibinfo{year}{2022}\natexlab{}.
\newblock \showarticletitle{Can OpenAI's Codex Fix Bugs?: An evaluation on
  QuixBugs}. In \bibinfo{booktitle}{\emph{3rd {IEEE/ACM} International Workshop
  on Automated Program Repair, APR@ICSE 2022, Pittsburgh, PA, USA, May 19,
  2022}}. \bibinfo{publisher}{{IEEE}}, \bibinfo{pages}{69--75}.
\newblock
\urldef\tempurl%
\url{https://doi.org/10.1145/3524459.3527351}
\showDOI{\tempurl}


\bibitem[{Rust Analyzer Team}(2020)]%
        {rustanalyzer}
\bibfield{author}{\bibinfo{person}{{Rust Analyzer Team}}.}
  \bibinfo{year}{2020}\natexlab{}.
\newblock \bibinfo{title}{Rust Analyzer}.
\newblock
  \bibinfo{howpublished}{\url{https://github.com/rust-lang/rust-analyzer}}.
\newblock


\bibitem[{Rust Team}(2023a)]%
        {rustlang}
\bibfield{author}{\bibinfo{person}{{Rust Team}}.}
  \bibinfo{year}{2023}\natexlab{a}.
\newblock \bibinfo{title}{Rust}.
\newblock \bibinfo{howpublished}{\url{https://www.rust-lang.org/}}.
\newblock


\bibitem[{Rust Team}(2023b)]%
        {rust-clippy}
\bibfield{author}{\bibinfo{person}{{Rust Team}}.}
  \bibinfo{year}{2023}\natexlab{b}.
\newblock \bibinfo{title}{Rust Clippy Static Analysis}.
\newblock \bibinfo{howpublished}{\url{https://doc.rust-lang.org/clippy}}.
\newblock


\bibitem[{Rust Team}(2023c)]%
        {rust-crates-io}
\bibfield{author}{\bibinfo{person}{{Rust Team}}.}
  \bibinfo{year}{2023}\natexlab{c}.
\newblock \bibinfo{title}{The Rust community's crate registry}.
\newblock \bibinfo{howpublished}{\url{https://crates.io}}.
\newblock


\bibitem[{Rust Team}(2023d)]%
        {rust-error-codes}
\bibfield{author}{\bibinfo{person}{{Rust Team}}.}
  \bibinfo{year}{2023}\natexlab{d}.
\newblock \bibinfo{title}{Rust error codes index}.
\newblock
  \bibinfo{howpublished}{\url{https://doc.rust-lang.org/error_codes/error-index.html}}.
\newblock


\bibitem[{Rust Team}(2023e)]%
        {rustsurvey}
\bibfield{author}{\bibinfo{person}{{Rust Team}}.}
  \bibinfo{year}{2023}\natexlab{e}.
\newblock \bibinfo{title}{Rust Survey}.
\newblock
  \bibinfo{howpublished}{\url{https://blog.rust-lang.org/2022/02/15/Rust-Survey-2021.html}}.
\newblock


\bibitem[Saparov and He(2023)]%
        {saparov2023language}
\bibfield{author}{\bibinfo{person}{Abulhair Saparov} {and} \bibinfo{person}{He
  He}.} \bibinfo{year}{2023}\natexlab{}.
\newblock \bibinfo{title}{Language Models Are Greedy Reasoners: A Systematic
  Formal Analysis of Chain-of-Thought}.
\newblock
\newblock
\showeprint[arxiv]{2210.01240}~[cs.CL]


\bibitem[Sintaha et~al\mbox{.}(2023)]%
        {katana}
\bibfield{author}{\bibinfo{person}{Mifta Sintaha}, \bibinfo{person}{Noor
  Nashid}, {and} \bibinfo{person}{Ali Mesbah}.}
  \bibinfo{year}{2023}\natexlab{}.
\newblock \showarticletitle{Katana: Dual Slicing Based Context for Learning Bug
  Fixes}.
\newblock \bibinfo{journal}{\emph{ACM Trans. Softw. Eng. Methodol.}}
  \bibinfo{volume}{32}, \bibinfo{number}{4}, Article \bibinfo{articleno}{100}
  (\bibinfo{date}{may} \bibinfo{year}{2023}), \bibinfo{numpages}{27}~pages.
\newblock
\showISSN{1049-331X}
\urldef\tempurl%
\url{https://doi.org/10.1145/3579640}
\showDOI{\tempurl}


\bibitem[{Stack Overflow}(2021)]%
        {sosurvey}
\bibfield{author}{\bibinfo{person}{{Stack Overflow}}.}
  \bibinfo{year}{2021}\natexlab{}.
\newblock \bibinfo{title}{Stack Overflow survey}.
\newblock
  \bibinfo{howpublished}{\url{https://insights.stackoverflow.com/survey/2021}}.
\newblock


\bibitem[Tan et~al\mbox{.}(2018)]%
        {DBLP:conf/icse/TanDGR18}
\bibfield{author}{\bibinfo{person}{Shin~Hwei Tan}, \bibinfo{person}{Zhen Dong},
  \bibinfo{person}{Xiang Gao}, {and} \bibinfo{person}{Abhik Roychoudhury}.}
  \bibinfo{year}{2018}\natexlab{}.
\newblock \showarticletitle{Repairing crashes in Android apps}. In
  \bibinfo{booktitle}{\emph{Proceedings of the 40th International Conference on
  Software Engineering, {ICSE} 2018, Gothenburg, Sweden, May 27 - June 03,
  2018}}, \bibfield{editor}{\bibinfo{person}{Michel Chaudron},
  \bibinfo{person}{Ivica Crnkovic}, \bibinfo{person}{Marsha Chechik}, {and}
  \bibinfo{person}{Mark Harman}} (Eds.). \bibinfo{publisher}{{ACM}},
  \bibinfo{pages}{187--198}.
\newblock
\urldef\tempurl%
\url{https://doi.org/10.1145/3180155.3180243}
\showDOI{\tempurl}


\bibitem[Tan and Roychoudhury(2015)]%
        {DBLP:conf/icse/TanR15}
\bibfield{author}{\bibinfo{person}{Shin~Hwei Tan} {and} \bibinfo{person}{Abhik
  Roychoudhury}.} \bibinfo{year}{2015}\natexlab{}.
\newblock \showarticletitle{relifix: Automated Repair of Software Regressions}.
  In \bibinfo{booktitle}{\emph{37th {IEEE/ACM} International Conference on
  Software Engineering, {ICSE} 2015, Florence, Italy, May 16-24, 2015, Volume
  1}}, \bibfield{editor}{\bibinfo{person}{Antonia Bertolino},
  \bibinfo{person}{Gerardo Canfora}, {and} \bibinfo{person}{Sebastian~G.
  Elbaum}} (Eds.). \bibinfo{publisher}{{IEEE} Computer Society},
  \bibinfo{pages}{471--482}.
\newblock
\urldef\tempurl%
\url{https://doi.org/10.1109/ICSE.2015.65}
\showDOI{\tempurl}


\bibitem[Touvron et~al\mbox{.}(2023a)]%
        {DBLP:journals/corr/abs-2302-13971}
\bibfield{author}{\bibinfo{person}{Hugo Touvron}, \bibinfo{person}{Thibaut
  Lavril}, \bibinfo{person}{Gautier Izacard}, \bibinfo{person}{Xavier
  Martinet}, \bibinfo{person}{Marie{-}Anne Lachaux},
  \bibinfo{person}{Timoth{\'{e}}e Lacroix}, {and} \bibinfo{person}{et al.}}
  \bibinfo{year}{2023}\natexlab{a}.
\newblock \showarticletitle{LLaMA: Open and Efficient Foundation Language
  Models}.
\newblock \bibinfo{journal}{\emph{CoRR}}  \bibinfo{volume}{abs/2302.13971}
  (\bibinfo{year}{2023}).
\newblock
\urldef\tempurl%
\url{https://doi.org/10.48550/arXiv.2302.13971}
\showDOI{\tempurl}
\showeprint[arXiv]{2302.13971}


\bibitem[Touvron et~al\mbox{.}(2023b)]%
        {DBLP:journals/corr/abs-2307-09288}
\bibfield{author}{\bibinfo{person}{Hugo Touvron}, \bibinfo{person}{Louis
  Martin}, \bibinfo{person}{Kevin Stone}, \bibinfo{person}{Peter Albert},
  \bibinfo{person}{Amjad Almahairi}, \bibinfo{person}{Yasmine Babaei}, {and}
  \bibinfo{person}{et al.}} \bibinfo{year}{2023}\natexlab{b}.
\newblock \showarticletitle{Llama 2: Open Foundation and Fine-Tuned Chat
  Models}.
\newblock \bibinfo{journal}{\emph{CoRR}}  \bibinfo{volume}{abs/2307.09288}
  (\bibinfo{year}{2023}).
\newblock
\urldef\tempurl%
\url{https://doi.org/10.48550/arXiv.2307.09288}
\showDOI{\tempurl}
\showeprint[arXiv]{2307.09288}


\bibitem[Tufano et~al\mbox{.}(2019)]%
        {DBLP:journals/tosem/TufanoWBPWP19}
\bibfield{author}{\bibinfo{person}{Michele Tufano}, \bibinfo{person}{Cody
  Watson}, \bibinfo{person}{Gabriele Bavota}, \bibinfo{person}{Massimiliano~Di
  Penta}, \bibinfo{person}{Martin White}, {and} \bibinfo{person}{Denys
  Poshyvanyk}.} \bibinfo{year}{2019}\natexlab{}.
\newblock \showarticletitle{An Empirical Study on Learning Bug-Fixing Patches
  in the Wild via Neural Machine Translation}.
\newblock \bibinfo{journal}{\emph{{ACM} Trans. Softw. Eng. Methodol.}}
  \bibinfo{volume}{28}, \bibinfo{number}{4} (\bibinfo{year}{2019}),
  \bibinfo{pages}{19:1--19:29}.
\newblock
\urldef\tempurl%
\url{https://doi.org/10.1145/3340544}
\showDOI{\tempurl}


\bibitem[van Tonder and Goues(2018)]%
        {DBLP:conf/icse/TonderG18}
\bibfield{author}{\bibinfo{person}{Rijnard van Tonder} {and}
  \bibinfo{person}{Claire~Le Goues}.} \bibinfo{year}{2018}\natexlab{}.
\newblock \showarticletitle{Static automated program repair for heap
  properties}. In \bibinfo{booktitle}{\emph{Proceedings of the 40th
  International Conference on Software Engineering, {ICSE} 2018, Gothenburg,
  Sweden, May 27 - June 03, 2018}}, \bibfield{editor}{\bibinfo{person}{Michel
  Chaudron}, \bibinfo{person}{Ivica Crnkovic}, \bibinfo{person}{Marsha
  Chechik}, {and} \bibinfo{person}{Mark Harman}} (Eds.).
  \bibinfo{publisher}{{ACM}}, \bibinfo{pages}{151--162}.
\newblock
\urldef\tempurl%
\url{https://doi.org/10.1145/3180155.3180250}
\showDOI{\tempurl}


\bibitem[Wang et~al\mbox{.}(2021)]%
        {DBLP:conf/emnlp/0034WJH21}
\bibfield{author}{\bibinfo{person}{Yue Wang}, \bibinfo{person}{Weishi Wang},
  \bibinfo{person}{Shafiq~R. Joty}, {and} \bibinfo{person}{Steven C.~H. Hoi}.}
  \bibinfo{year}{2021}\natexlab{}.
\newblock \showarticletitle{CodeT5: Identifier-aware Unified Pre-trained
  Encoder-Decoder Models for Code Understanding and Generation}. In
  \bibinfo{booktitle}{\emph{Proceedings of the 2021 Conference on Empirical
  Methods in Natural Language Processing, {EMNLP} 2021, Virtual Event / Punta
  Cana, Dominican Republic, 7-11 November, 2021}},
  \bibfield{editor}{\bibinfo{person}{Marie{-}Francine Moens},
  \bibinfo{person}{Xuanjing Huang}, \bibinfo{person}{Lucia Specia}, {and}
  \bibinfo{person}{Scott~Wen{-}tau Yih}} (Eds.).
  \bibinfo{publisher}{Association for Computational Linguistics},
  \bibinfo{pages}{8696--8708}.
\newblock
\urldef\tempurl%
\url{https://doi.org/10.18653/v1/2021.emnlp-main.685}
\showDOI{\tempurl}


\bibitem[Wen et~al\mbox{.}(2018)]%
        {DBLP:conf/icse/WenCWHC18}
\bibfield{author}{\bibinfo{person}{Ming Wen}, \bibinfo{person}{Junjie Chen},
  \bibinfo{person}{Rongxin Wu}, \bibinfo{person}{Dan Hao}, {and}
  \bibinfo{person}{Shing{-}Chi Cheung}.} \bibinfo{year}{2018}\natexlab{}.
\newblock \showarticletitle{Context-aware patch generation for better automated
  program repair}. In \bibinfo{booktitle}{\emph{Proceedings of the 40th
  International Conference on Software Engineering, {ICSE} 2018, Gothenburg,
  Sweden, May 27 - June 03, 2018}}, \bibfield{editor}{\bibinfo{person}{Michel
  Chaudron}, \bibinfo{person}{Ivica Crnkovic}, \bibinfo{person}{Marsha
  Chechik}, {and} \bibinfo{person}{Mark Harman}} (Eds.).
  \bibinfo{publisher}{{ACM}}, \bibinfo{pages}{1--11}.
\newblock
\urldef\tempurl%
\url{https://doi.org/10.1145/3180155.3180233}
\showDOI{\tempurl}


\bibitem[White et~al\mbox{.}(2016)]%
        {DBLP:conf/kbse/WhiteTVP16}
\bibfield{author}{\bibinfo{person}{Martin White}, \bibinfo{person}{Michele
  Tufano}, \bibinfo{person}{Christopher Vendome}, {and} \bibinfo{person}{Denys
  Poshyvanyk}.} \bibinfo{year}{2016}\natexlab{}.
\newblock \showarticletitle{Deep learning code fragments for code clone
  detection}. In \bibinfo{booktitle}{\emph{Proceedings of the 31st {IEEE/ACM}
  International Conference on Automated Software Engineering, {ASE} 2016,
  Singapore, September 3-7, 2016}}, \bibfield{editor}{\bibinfo{person}{David
  Lo}, \bibinfo{person}{Sven Apel}, {and} \bibinfo{person}{Sarfraz Khurshid}}
  (Eds.). \bibinfo{publisher}{{ACM}}, \bibinfo{pages}{87--98}.
\newblock
\urldef\tempurl%
\url{https://doi.org/10.1145/2970276.2970326}
\showDOI{\tempurl}


\bibitem[Xia et~al\mbox{.}(2023a)]%
        {fitrepair}
\bibfield{author}{\bibinfo{person}{Chunqiu~Steven Xia}, \bibinfo{person}{Yifeng
  Ding}, {and} \bibinfo{person}{Lingming Zhang}.}
  \bibinfo{year}{2023}\natexlab{a}.
\newblock \bibinfo{title}{Revisiting the Plastic Surgery Hypothesis via Large
  Language Models}.
\newblock
\newblock
\showeprint[arxiv]{2303.10494}~[cs.SE]


\bibitem[Xia et~al\mbox{.}(2022)]%
        {xia2022practical}
\bibfield{author}{\bibinfo{person}{Chunqiu~Steven Xia},
  \bibinfo{person}{Yuxiang Wei}, {and} \bibinfo{person}{Lingming Zhang}.}
  \bibinfo{year}{2022}\natexlab{}.
\newblock \bibinfo{title}{Practical Program Repair in the Era of Large
  Pre-trained Language Models}.
\newblock
\newblock
\showeprint[arxiv]{2210.14179}~[cs.SE]


\bibitem[Xia et~al\mbox{.}(2023b)]%
        {DBLP:conf/icse/XiaWZ23}
\bibfield{author}{\bibinfo{person}{Chunqiu~Steven Xia},
  \bibinfo{person}{Yuxiang Wei}, {and} \bibinfo{person}{Lingming Zhang}.}
  \bibinfo{year}{2023}\natexlab{b}.
\newblock \showarticletitle{Automated Program Repair in the Era of Large
  Pre-trained Language Models}. In \bibinfo{booktitle}{\emph{45th {IEEE/ACM}
  International Conference on Software Engineering, {ICSE} 2023, Melbourne,
  Australia, May 14-20, 2023}}. \bibinfo{publisher}{{IEEE}},
  \bibinfo{pages}{1482--1494}.
\newblock
\urldef\tempurl%
\url{https://doi.org/10.1109/ICSE48619.2023.00129}
\showDOI{\tempurl}


\bibitem[Xuan et~al\mbox{.}(2017)]%
        {DBLP:journals/tse/XuanMDCMDBM17}
\bibfield{author}{\bibinfo{person}{Jifeng Xuan}, \bibinfo{person}{Matias
  Martinez}, \bibinfo{person}{Favio Demarco}, \bibinfo{person}{Maxime Clement},
  \bibinfo{person}{Sebastian R.~Lamelas Marcote}, \bibinfo{person}{Thomas
  Durieux}, \bibinfo{person}{Daniel~Le Berre}, {and} \bibinfo{person}{Martin
  Monperrus}.} \bibinfo{year}{2017}\natexlab{}.
\newblock \showarticletitle{Nopol: Automatic Repair of Conditional Statement
  Bugs in Java Programs}.
\newblock \bibinfo{journal}{\emph{{IEEE} Trans. Software Eng.}}
  \bibinfo{volume}{43}, \bibinfo{number}{1} (\bibinfo{year}{2017}),
  \bibinfo{pages}{34--55}.
\newblock
\urldef\tempurl%
\url{https://doi.org/10.1109/TSE.2016.2560811}
\showDOI{\tempurl}


\bibitem[Yang et~al\mbox{.}(2023)]%
        {transplanfix}
\bibfield{author}{\bibinfo{person}{Deheng Yang}, \bibinfo{person}{Xiaoguang
  Mao}, \bibinfo{person}{Liqian Chen}, \bibinfo{person}{Xuezheng Xu},
  \bibinfo{person}{Yan Lei}, \bibinfo{person}{David Lo}, {and}
  \bibinfo{person}{Jiayu He}.} \bibinfo{year}{2023}\natexlab{}.
\newblock \showarticletitle{TransplantFix: Graph Differencing-Based Code
  Transplantation for Automated Program Repair}. In
  \bibinfo{booktitle}{\emph{Proceedings of the 37th IEEE/ACM International
  Conference on Automated Software Engineering}} (Rochester, MI, USA)
  \emph{(\bibinfo{series}{ASE '22})}. \bibinfo{publisher}{Association for
  Computing Machinery}, \bibinfo{address}{New York, NY, USA}, Article
  \bibinfo{articleno}{107}, \bibinfo{numpages}{13}~pages.
\newblock
\showISBNx{9781450394758}
\urldef\tempurl%
\url{https://doi.org/10.1145/3551349.3556893}
\showDOI{\tempurl}


\bibitem[Ye et~al\mbox{.}(2022)]%
        {DBLP:conf/icse/YeMM22}
\bibfield{author}{\bibinfo{person}{He Ye}, \bibinfo{person}{Matias Martinez},
  {and} \bibinfo{person}{Martin Monperrus}.} \bibinfo{year}{2022}\natexlab{}.
\newblock \showarticletitle{Neural Program Repair with Execution-based
  Backpropagation}. In \bibinfo{booktitle}{\emph{44th {IEEE/ACM} 44th
  International Conference on Software Engineering, {ICSE} 2022, Pittsburgh,
  PA, USA, May 25-27, 2022}}. \bibinfo{publisher}{{ACM}},
  \bibinfo{pages}{1506--1518}.
\newblock
\urldef\tempurl%
\url{https://doi.org/10.1145/3510003.3510222}
\showDOI{\tempurl}


\bibitem[Yuan and Banzhaf(2020)]%
        {DBLP:journals/tse/YuanB20}
\bibfield{author}{\bibinfo{person}{Yuan Yuan} {and} \bibinfo{person}{Wolfgang
  Banzhaf}.} \bibinfo{year}{2020}\natexlab{}.
\newblock \showarticletitle{{ARJA:} Automated Repair of Java Programs via
  Multi-Objective Genetic Programming}.
\newblock \bibinfo{journal}{\emph{{IEEE} Trans. Software Eng.}}
  \bibinfo{volume}{46}, \bibinfo{number}{10} (\bibinfo{year}{2020}),
  \bibinfo{pages}{1040--1067}.
\newblock
\urldef\tempurl%
\url{https://doi.org/10.1109/TSE.2018.2874648}
\showDOI{\tempurl}


\bibitem[Zhu et~al\mbox{.}(2021)]%
        {DBLP:conf/sigsoft/ZhuSXZY0Z21}
\bibfield{author}{\bibinfo{person}{Qihao Zhu}, \bibinfo{person}{Zeyu Sun},
  \bibinfo{person}{Yuan{-}an Xiao}, \bibinfo{person}{Wenjie Zhang},
  \bibinfo{person}{Kang Yuan}, \bibinfo{person}{Yingfei Xiong}, {and}
  \bibinfo{person}{Lu Zhang}.} \bibinfo{year}{2021}\natexlab{}.
\newblock \showarticletitle{A syntax-guided edit decoder for neural program
  repair}. In \bibinfo{booktitle}{\emph{{ESEC/FSE} '21: 29th {ACM} Joint
  European Software Engineering Conference and Symposium on the Foundations of
  Software Engineering, Athens, Greece, August 23-28, 2021}},
  \bibfield{editor}{\bibinfo{person}{Diomidis Spinellis},
  \bibinfo{person}{Georgios Gousios}, \bibinfo{person}{Marsha Chechik}, {and}
  \bibinfo{person}{Massimiliano~Di Penta}} (Eds.). \bibinfo{publisher}{{ACM}},
  \bibinfo{pages}{341--353}.
\newblock
\urldef\tempurl%
\url{https://doi.org/10.1145/3468264.3468544}
\showDOI{\tempurl}


\end{thebibliography}
